%% file: draft_arxiv.tex
\documentclass[aps,prl,twocolumn,superscriptaddress]{revtex4-2}
\input{preamble}

\begin{document}

\title{Oscillatory-nonnormal decomposition of dissipation in Ornstein--Uhlenbeck processes}

\author{Ryuna Nagayama}
\email{ryuna.nagayama@ubi.s.u-tokyo.ac.jp}
\affiliation{Department of Physics, Graduate School of Science, The University of Tokyo, 7-3-1 Hongo, Bunkyo-ku, Tokyo 113-0033, Japan}
\author{Artemy Kolchinsky}
\affiliation{ICREA-Complex Systems Lab, Pompeu Fabra University, 08003 Barcelona, Spain}
\affiliation{Universal Biology Institute, Graduate School of Science, The University of Tokyo, 7-3-1 Hongo, Bunkyo-ku, Tokyo 113-0033, Japan}
\author{Sosuke Ito}
\affiliation{Department of Physics, Graduate School of Science, The University of Tokyo, 7-3-1 Hongo, Bunkyo-ku, Tokyo 113-0033, Japan}
\affiliation{Universal Biology Institute, Graduate School of Science, The University of Tokyo, 7-3-1 Hongo, Bunkyo-ku, Tokyo 113-0033, Japan}

\date{\today}

\begin{abstract}
We provide a decomposition of the steady-state entropy production rate associated with an Ornstein--Uhlenbeck process into two contributions: one associated with oscillatory behavior and one associated with nonnormality. We also show that each contribution is associated with a different fundamental trade-off. The oscillatory contribution leads to the dissipation-coherence trade-off for noise-induced oscillations, which bounds the entropy production per oscillatory period by the number of oscillations within one correlation time. Notably, the trade-off is twice as strict as those conjectured or derived for other systems.
The nonnormal contribution leads to a trade-off between entropy production and acceleration of relaxation.
We also demonstrate the decomposition using a simple bead-spring model.
\end{abstract}

\maketitle

\textit{Introduction}.---%
The Ornstein--Uhlenbeck (OU) process~\cite{uhlenbeck1930theory,wang1945theory} describes the statistics of systems with linear Langevin dynamics~\cite{risken1989fokker,gardiner2009stochastic,lax1960fluctuations}. Well-known examples include driven colloidal particles in a harmonic potential~\cite{schmiedl2008efficiency,blickle2012realization}, electrical circuits~\cite{van2004power,ciliberto2013heat,chiang2017electrical,melanson2025thermodynamic}, bead-spring systems~\cite{mura2018nonequilibrium,li2019quantifying,van2020entropy}, and fluctuations around a stable fixed point~\cite{van1992stochastic,kubo1966fluctuation,kwon2005structure,hasselmann1976stochastic,aslyamov2026macroscopic}.
In nonequilibrium OU processes with broken detailed balance, the probability flux exhibits ongoing circulation~\cite{tomita1974irreversible}. This can induce persistent oscillations~\cite{thomas2019phase,mckane2005predator,mckane2007amplified,westermark2009quantification,wallace2011emergent,gilson2023entropy,sekizawa2024decomposing,nartallo2026,dinis2012fluctuation,alonso2007stochastic,kuske2007sustained,lugo2008quasicycles,mura2018nonequilibrium}, accelerate the system's relaxation~\cite{hwang1993accelerating,lelievre2013optimal}, and give rise to transient amplification~\cite{weiss2003coordinate,fyodorov2025nonorthogonal,nartallo2024broken,sornette2025life,troude2026pseudo}.

Recent developments in stochastic thermodynamics have revealed quantitative relationships between such nonequilibrium phenomena and the associated entropy production (EP)~\cite{sekimoto2010stochastic,seifert2012stochastic,Shiraishi2023,falasco2025macroscopic}. 
However, several fundamental questions remain open, even in the simple and ubiquitous case of OU processes. One of these questions concerns the relationship between nonequilibrium and spectral properties. Although it is understood that complex-valued eigenvalues are a signature of broken detailed balance, the precise relationship between thermodynamics and eigenvalue localization remains under investigation~\cite{barato2017coherence,uhl2019affinity,ohga2023thermodynamic,xu2025thermodynamic,kolchinsky2026cycle,qian2000pumped,nguyen2018phase,oberreiter2019subharmonic,marsland2019thermodynamic,del2020robust,del2020high,oberreiter2021stochastic,remlein2022coherence,shiraishi2023entropy,kolchinsky2024thermodynamic,pietzonka2024thermodynamic,gao2024thermodynamic,zheng2024topological}. This relationship has implications for various biological and artificial systems, since complex eigenvalues are necessary for resonant response to periodic driving~\cite{hanggi1982stochastic} and coherent stochastic oscillations~\cite{barato2017coherence}. In fact, it has been conjectured that there is a fundamental \emph{dissipation-coherence trade-off} (DCT) that specifies the minimum EP required to maintain coherent oscillations~\cite{oberreiter2022universal}. However, until now, this trade-off has only been proved in special cases such as nonlinear oscillators subject to weak noise~\cite{santolin2025dissipation,nagayama2025duality,kolchinsky2025comment}. Establishing the DCT for noise-induced oscillations, as described by an OU process, remains a critical challenge.

Another open question concerns the relationship between nonequilibrium and nonnormality~\cite{trefethen2020spectra}. Nonnormality occurs when a system's generator does not commute with its adjoint, and it is linked to transient phenomena such as amplification~\cite{farrell1994variance,penland1995optimal,hennequin2012non,biancalani2017giant}, strong response to perturbations~\cite{farrell1996generalized,ahmadian2015properties}, and temporal synchronization~\cite{troude2026pseudo}. Although nonnormality is known to be a signature of nonequilibrium~\cite{weiss2003coordinate,polettini2013nonconvexity,nartallo2024broken,fyodorov2025nonorthogonal}, its quantitative relationship with thermodynamic driving has been largely unexplored. Recently, it has been suggested that nonnormality increases EP in OU processes~\cite{fyodorov2025nonorthogonal}. However, to our knowledge, analytical evidence for such an increase remains limited to some two-dimensional models~\cite{sornette2025life,troude2026pseudo} and the asymptotic behavior of a high-dimensional random model~\cite{fyodorov2025nonorthogonal}; a proof for general systems is still lacking.

In this Letter, we address these fundamental issues by establishing an exact decomposition of the steady-state EP rate (EPR) in OU processes into two contributions: one from oscillatory behavior and one from nonnormality. 
This decomposition provides analytical evidence for the increase in dissipation due to the system's nonnormality. 
As an application of the decomposition, we demonstrate that the oscillatory contribution obeys the DCT. Moreover, we show that the DCT is twice as strict in the OU process as in nonlinear systems, pointing to the intrinsic inefficiency of noise-induced oscillations. 
As another application, we use the decomposition to reveal that acceleration of relaxation requires nonnormality, and thus contributes to the nonnormal EPR. 
Finally, we demonstrate the decomposition numerically using a simple toy model.

\textit{Setup}.---%
We consider an $N$-dimensional system described by a multidimensional OU process~\cite{uhlenbeck1930theory,wang1945theory}. At time $t$, the probability distribution $p_t(\bm{x})$ of the system's state evolves according to the Fokker--Planck equation~\cite{risken1989fokker,gardiner2009stochastic},
\begin{align}
    \partial_t p_t(\bm{x})=\mathcal{L}[p_t(\bm{x})]\coloneqq\bm{\nabla}\cdot(p_t(\bm{x})\mathsf{K}\bm{x}+\mathsf{D}\bm{\nabla}p_t(\bm{x}))  \,,
    \label{Fokker--Planck eq}
\end{align}
with \textit{drift matrix} $\mathsf{K}$ and 
positive-definite \textit{diffusion matrix} $\mathsf{D}$. Here, $\mathcal{L}$ is the generator of time evolution. To ensure the system's stability, we assume that all eigenvalues of $\mathsf{K}$ have positive real parts. The Fokker--Planck equation~\eqref{Fokker--Planck eq} can be written as a continuity equation $\partial_tp_t(\bm{x})=-\bm{\nabla}\cdot\bm{j}(\bm{x})$ with flux field $\bm{j}(\bm{x})\coloneqq-p_t(\bm{x})\mathsf{K}\bm{x}-\mathsf{D}\bm{\nabla}p_t(\bm{x})$.

In the following, we focus on the system's steady state. Standard results~\cite{risken1989fokker} show that the stationary distribution $p^{\mathrm{st}}(\bm{x})$ is a Gaussian whose covariance matrix $\mathsf{V}$ solves the continuous-time Lyapunov equation $\mathsf{K}\mathsf{V}+\mathsf{V}\mathsf{K}^{\top}-2\mathsf{D}=\mathsf{0}$~\cite{risken1989fokker}.
Here, $\mathsf{0}$ is the zero matrix, and $\top$ denotes transpose.

In steady state, the flux field is given by $\bm{j}^{\mathrm{st}}(\bm{x})\coloneqq-p^{\mathrm{st}}(\bm{x})\mathsf{K}\bm{x}-\mathsf{D}\bm{\nabla}p^{\mathrm{st}}(\bm{x})$, which is divergence-free.
The steady state is called the equilibrium state when $\bm{j}^{\mathrm{st}}(\bm{x})$ vanishes everywhere, which is also called the condition of \textit{detailed balance}. Conversely, the steady state is nonequilibrium if $\bm{j}^{\mathrm{st}}(\bm{x})$ does not vanish everywhere.
In the following, we only consider the case where $N\geq2$, since one-dimensional systems always relax to equilibrium states.

The most common measure of nonequilibrium is the EPR. In the steady state, it is defined as $\sigma^{\mathrm{st}}\coloneqq\int d\bm{x}\,\bm{j}^{\mathrm{st}}(\bm{x})^{\top}(p^{\mathrm{st}}(\bm{x})\mathsf{D})^{-1}\bm{j}^{\mathrm{st}}(\bm{x})$~\cite{seifert2012stochastic}. 
In general, the EPR quantifies the degree of statistical irreversibility exhibited by the dynamics. In microscopic systems that obey the principle of local detailed balance~\cite{kondepudi2014modern,beard2007relationship,maes2021local}, it can also be understood as the rate of production of thermodynamic entropy.

\textit{Canonical coordinate system}.---%
To clearly characterize nonequilibrium, we consider a whitened coordinate system $\bm{x}\mapsto\tilde{\bm{x}}\coloneqq\mathsf{V}^{-1/2}\bm{x}$, which makes the steady-state covariance matrix the identity matrix $\mathsf{I}$. 
In these coordinates, the system is described by a Fokker--Planck equation with transformed drift and diffusion matrices
\begin{align}
    \tilde{\mathsf{K}}\coloneqq\mathsf{V}^{-\frac{1}{2}}\mathsf{K}\mathsf{V}^{\frac{1}{2}},\;
    \tilde{\mathsf{D}}\coloneqq\mathsf{V}^{-\frac{1}{2}}\mathsf{D}\mathsf{V}^{-\frac{1}{2}}.
    \label{drift matrix}
\end{align}
Using the Lyapunov equation, we may write the symmetric and antisymmetric parts of $\tilde{\mathsf{K}}$ as $\tilde{\mathsf{K}}_{+}\coloneqq(\tilde{\mathsf{K}}+\tilde{\mathsf{K}}^{\top})/2=\mathsf{V}^{-1/2}\mathsf{D}\mathsf{V}^{-1/2}=\tilde{\mathsf{D}}$ and $\tilde{\mathsf{K}}_{-}\coloneqq(\tilde{\mathsf{K}}-\tilde{\mathsf{K}}^{\top})/2  =\mathsf{V}^{-1/2}(\mathsf{K}\mathsf{V}-\mathsf{D})\mathsf{V}^{-1/2}$, respectively. As shown in Appendix~A, these relations and the system's linearity~\cite{landi2013entropy,godreche2018characterising} enable us to express the steady-state EPR as
\begin{align}
    \sigma^{\mathrm{st}}=\mathrm{tr}(\tilde{\mathsf{K}}_{-}^{\top}\tilde{\mathsf{K}}_{+}^{-1}\tilde{\mathsf{K}}_{-}).
    \label{EPR simple form}
\end{align}
Thus, in the whitened coordinate system, the EPR quantifies the relative magnitude of the asymmetry of $\tilde{\mathsf{K}}$.

Equilibrium (i.e., detailed balance) is equivalent to self-adjoint symmetry $\tilde{\mathsf{K}}=\tilde{\mathsf{K}}^{\top}$, so that $\tilde{\mathsf{K}}_{-}=\mathsf{0}$. Note that this condition is different from the system's reciprocity $\mathsf{K}=\mathsf{K}^{\top}$. Indeed, a nonreciprocal system can be detailed balanced if it is in contact with multiple heat bath~\cite{loos2020irreversibility}.
Mathematically, this symmetry $\tilde{\mathsf{K}}=\tilde{\mathsf{K}}^{\top}$ holds if and only if $\tilde{\mathsf{K}}$ has only real eigenvalues and is \textit{normal}, meaning that it commutes with its own transpose, i.e., $\tilde{\mathsf{K}}\tilde{\mathsf{K}}^{\top}=\tilde{\mathsf{K}}^{\top}\tilde{\mathsf{K}}$. Therefore, detailed balance may be broken in either or both of two ways: $\tilde{\mathsf{K}}$ may have complex eigenvalues and/or be nonnormal.

There is a direct correspondence between the eigenvalues of $\tilde{\mathsf{K}}$ and those of the generator $\mathcal{L}$ in Eq.~\eqref{Fokker--Planck eq}. Let $\{\lambda_n\}_{n=1}^N$ denote the eigenvalues of $\tilde{\mathsf{K}}$, repeated according to their algebraic multiplicities and labeled so that $\mathrm{Re}(\lambda_1)\leq\mathrm{Re}(\lambda_2)\leq\cdots\leq\mathrm{Re}(\lambda_N)$.
Here and in the following, the real and imaginary parts of a complex number are denoted by $\mathrm{Re}(\cdots)$ and $\mathrm{Im}(\cdots)$, respectively. 
Since $\tilde{\mathsf{K}}$ is similar to $\mathsf{K}$, $\{\lambda_n\}_{n=1}^N$ are also eigenvalues of $\mathsf{K}$, and their real parts are positive. The spectrum of $\mathcal{L}$ is given by superpositions of the eigenvalues of $-\mathsf{K}$ with nonnegative integer coefficients, $\{-\sum_{k=1}^Nn_k\lambda_k\mid n_k\in\mathbb{Z}_{\geq0}\}$~\cite{metafune2002spectrum,leen2016eigenfunctions,sekizawa2025koopman}. Therefore, $\tilde{\mathsf{K}}$ has complex eigenvalues if and only if $\mathcal{L}$ has complex eigenvalues.

Although the nonnormality of the drift matrix depends on the coordinate system~\cite{weiss2003coordinate}, we can regard the nonnormality of $\tilde{\mathsf{K}}$ as intrinsic. This is because the nonnormality of $\tilde{\mathsf{K}}$ 
is equivalent to that of the twisted generator 
$\tilde{\mathcal{L}}[\psi(\bm{x})]\coloneqq p^{\mathrm{st}}(\bm{x})^{-1/2}\mathcal{L}[p^{\mathrm{st}}(\bm{x})^{1/2}\psi(\bm{x})]$~\cite{polettini2013nonconvexity}, as shown in Supplemental Material (SM)~\cite{SuppMat}. 
The nonnormality of $\tilde{\mathsf{K}}$ is also equivalent to that of $\mathsf{K}_{D}\coloneqq\mathsf{D}^{-1/2}\mathsf{K}\mathsf{D}^{1/2}$ (see Appendix~B), where $\mathsf{K}_{D}$ is the drift matrix in the coordinate system in which the diffusion matrix becomes the identity. This coordinate system has been used to consider the system's nonnormality~\cite{weiss2003coordinate,fyodorov2025nonorthogonal}.

\begin{figure}
    \centering
    \includegraphics[width=\linewidth]{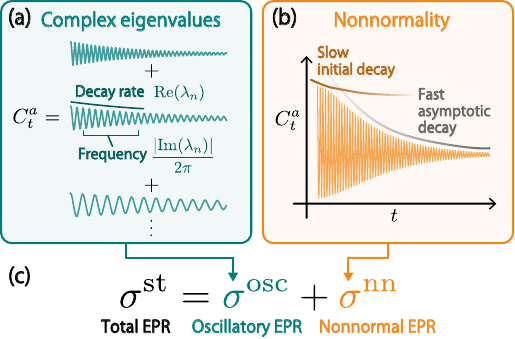}
    \caption{(a) The autocorrelation $C_t^a$ may be expanded into a superposition of damped oscillatory modes, whose decay rates and frequencies are determined by the eigenvalues of $\tilde{\mathsf{K}}$. Oscillatory behavior emerges only if $\tilde{\mathsf{K}}$ possesses complex eigenvalues. (b) The nonnormality of $\tilde{\mathsf{K}}$ allows the short-time decay rate of $C_t^a$ to be smaller than its asymptotic decay rate in the long-time regime. (c) We establish a decomposition of the steady-state EPR into the oscillatory contribution and the nonnormal contribution [Eq.~\eqref{oscillatory-nonnormal decomposition}].}
    \label{fig:schematics}
\end{figure}

As shown in Figs.~\ref{fig:schematics}(a) and ~\ref{fig:schematics}(b), complex eigenvalues and nonnormality each have distinct effects on the autocorrelation function of generic linear observables (see Appendix~C for details). Here, the autocorrelation function of a linear observable $a(\tilde{\bm{x}})\coloneqq\bm{a}^{\top}\tilde{\bm{x}}$ with $\bm{a}\in\mathbb{R}^N$ is defined as $C^a_t\coloneqq\langle a(\tilde{\bm{x}}_t)a(\tilde{\bm{x}}_0)\rangle_{\mathrm{st}}$, where $\langle\cdots\rangle_{\mathrm{st}}$ is the expected value in the steady state.

\textit{Oscillatory-nonnormal decomposition of EPR.---}
We now present our main result: the oscillatory-nonnormal decomposition [Fig.~\ref{fig:schematics}(c)]. A key step is the use of the \emph{Schur decomposition} of $\tilde{\mathsf{K}}$~\cite{Horn1985}. This allows us to express $\tilde{\mathsf{K}}=\mathsf{U}\mathsf{T}\mathsf{U}^{\dag}$, where $\mathsf{U}$ is a unitary matrix, $\mathsf{T}$ is an upper triangular matrix with diagonal entries $T_{nn}=\lambda_n$, and $\dag$ denotes the conjugate transpose. For normal $\tilde{\mathsf{K}}$, $\mathsf{T}$ becomes diagonal, and the Schur decomposition is the same as the eigendecomposition. For nonnormal $\tilde{\mathsf{K}}$, $\mathsf{T}$ is no longer diagonal, and the Schur decomposition may be understood as a generalization of an orthonormal eigendecomposition. Importantly, unlike the eigendecomposition, every square matrix has a Schur decomposition.

We rewrite Eq.~\eqref{EPR simple form} using the Schur decomposition. We define the Hermitian and skew-Hermitian parts of $\mathsf{T}$ as $\mathsf{H}\coloneqq(\mathsf{T}+\mathsf{T}^{\dag})/2$ and $\mathsf{S}\coloneqq(\mathsf{T}-\mathsf{T}^{\dag})/2$, respectively. Their diagonal elements correspond to the eigenvalues of $\tilde{\mathsf{K}}$ as $H_{nn}=\mathrm{Re}(\lambda_n)$ and $S_{nn}=\mathrm{i}\mathrm{Im}(\lambda_n)$.
Since $\tilde{\mathsf{K}}$ is real and satisfies $\tilde{\mathsf{K}}^{\top}=\tilde{\mathsf{K}}^{\dag}$, these matrices satisfy $\tilde{\mathsf{K}}_{+}=\mathsf{U}\mathsf{H}\mathsf{U}^{\dag}$, $\tilde{\mathsf{K}}_{-}=\mathsf{U}\mathsf{S}\mathsf{U}^{\dag}$, and $\tilde{\mathsf{K}}_{-}^{\top}=\tilde{\mathsf{K}}_{-}^{\dag}=\mathsf{U}\mathsf{S}^{\dag}\mathsf{U}^{\dag}$.
These relations rewrite Eq.~\eqref{EPR simple form} as 
\begin{align}
    \sigma^{\mathrm{st}}=\mathrm{tr}(\mathsf{S}^{\dag}\mathsf{H}^{-1}\mathsf{S})=\|\mathsf{S}\|^2_{\mathsf{H}^{-1}},
    \label{EPR Schur form}
\end{align}
where the norm $\|\cdot\|_{\mathsf{H}^{-1}}$ is induced by the inner product $\langle\mathsf{M},\mathsf{M}'\rangle_{\mathsf{H}^{-1}}\coloneqq\mathrm{tr}(\mathsf{M}^{\dag}\mathsf{H}^{-1}\mathsf{M}')$. Observe that  $\mathsf{H}=\mathsf{U}^{\dag}\tilde{\mathsf{K}}_{+}\mathsf{U}=\mathsf{U}^{\dag}\tilde{\mathsf{D}}\mathsf{U}$ is positive definite.

Next, we introduce the subspace $\mathsf{H}\mathcal{D}\coloneqq\{\mathsf{H}\mathsf{X}\mid\mathsf{X}\in\mathcal{D}\}$, where $\mathcal{D}$ denotes the space of $N\times N$ complex-valued diagonal matrices. This subspace characterizes the system's normality. Indeed, $\tilde{\mathsf{K}}$ is normal if and only if  $\mathsf{S}\in\mathsf{H}\mathcal{D}$ as shown in Appendix~D. Thus, the nonnormality of $\tilde{\mathsf{K}}$ can be measured as the distance between $\mathsf{S}$ and $\mathsf{H}\mathcal{D}$.

Using the projection of $\mathsf{S}$ onto $\mathsf{H}\mathcal{D}$,
\begin{align}
    \mathsf{S}^{\star}\coloneqq\argmin_{\mathsf{S}'\in\mathsf{H}\mathcal{D}}\|\mathsf{S}-\mathsf{S}'\|^2_{\mathsf{H}^{-1}},
    \label{projection}
\end{align}
and the Pythagorean Theorem $\|\mathsf{S}\|^2_{\mathsf{H}^{-1}}=\|\mathsf{S}^{\star}\|^2_{\mathsf{H}^{-1}}+\|\mathsf{S}-\mathsf{S}^{\star}\|^2_{\mathsf{H}^{-1}}$, we obtain the oscillatory-nonnormal decomposition of the steady-state EPR:
\begin{align}
    \sigma^{\mathrm{st}}=\sigma^{\mathrm{osc}}+\sigma^{\mathrm{nn}}.
    \label{oscillatory-nonnormal decomposition}
\end{align}
Here, we define the oscillatory EPR as
\begin{align}
    \sigma^{\mathrm{osc}}\coloneqq\|\mathsf{S}^{\star}\|^2_{\mathsf{H}^{-1}}=\sum_{n=1}^N\frac{\mathrm{Im}(\lambda_n)^2}{\mathrm{Re}(\lambda_n)},
    \label{oscillatory EPR}
\end{align}
where the last equality is obtained by solving the projection as $\mathsf{S}^{\star}=\mathsf{H}\mathsf{X}^{\star}$ with $X^{\star}_{nm}=\mathrm{i}\delta_{nm}\mathrm{Im}(\lambda_n)/\mathrm{Re}(\lambda_n)$ (see Appendix~D for details). This EPR is determined solely by the eigenvalues of $\tilde{\mathsf{K}}$ and measures the total intensity of the damped oscillatory eigenmodes: Eq.~\eqref{oscillatory EPR} is the sum of the modes' squared angular frequencies $\mathrm{Im}(\lambda_n)^2$ weighted by their decay times $1/\mathrm{Re}(\lambda_n)$.
We also define the nonnormal EPR as
\begin{align}
    \sigma^{\mathrm{nn}}\coloneqq\|\mathsf{S}-\mathsf{S}^{\star}\|^2_{\mathsf{H}^{-1}}.
    \label{nonnormal EPR}
\end{align}
This is a squared distance between $\mathsf{S}$ and $\mathsf{H}\mathcal{D}$, and thus measures the nonnormality of $\tilde{\mathsf{K}}$. Indeed, $\sigma^{\mathrm{nn}}$ is nonnegative and vanishes if and only if $\tilde{\mathsf{K}}$ is normal.
We note that the derivation based on projection implies that the oscillatory-nonnormal decomposition is a type of \textit{geometric decomposition}, which has been used to decompose the EPR of various systems into multiple nonnegative contributions~\cite{ito2020unified,dechant2022geometric_R,dechant2022geometric_E,ito2023geometric,yoshimura2023housekeeping,kobayashi2022hessian,kolchinsky2026generalized,aguilera2026inferring,nagayama2023geometric,yoshimura2024two,yoshimura2025force,sekizawa2024decomposing}.

The oscillatory-nonnormal decomposition implies that, in the steady state of the OU process, dissipation arises separately from the complex eigenvalues and the nonnormality of $\tilde{\mathsf{K}}$. Consequently, steady states can be systematically classified into four distinct types depending on whether $\tilde{\mathsf{K}}$ has complex eigenvalues and whether $\tilde{\mathsf{K}}$ is nonnormal. The first type is an equilibrium state $(\sigma^{\mathrm{osc}}=0,\sigma^{\mathrm{nn}}=0)$, in which the system satisfies detailed balance. The second type is a purely oscillatory steady state $(\sigma^{\mathrm{osc}}>0,\sigma^{\mathrm{nn}}=0)$, where the total EPR is given by $\sigma^{\mathrm{st}}=\sum_{n=1}^N\mathrm{Im}(\lambda_n)^2/\mathrm{Re}(\lambda_n)$. This expression of $\sigma^{\mathrm{st}}$ was previously obtained for two special cases: systems with cyclic symmetry~\cite{godreche2018characterising} and the stochastic Amari neural field model~\cite{lucente2025entropy}.
The third type is a purely nonnormal steady state $(\sigma^{\mathrm{osc}}=0,\sigma^{\mathrm{nn}}>0)$. A system driven by a conservative force and in contact with multiple heat baths is a typical example of this type. Indeed, in such a system, $\tilde{\mathsf{K}}$ cannot have complex eigenvalues because $\mathsf{K}$ is the Hessian of the quadratic potential, and thus symmetric. 
The fourth type is an oscillatory-nonnormal steady state $(\sigma^{\mathrm{osc}}>0,\sigma^{\mathrm{nn}}>0)$. As we will show with a numerical example, this scenario can arise from the coupling of multiple oscillators.

Since the normality of $\tilde{\mathsf{K}}$ is equivalent to that of $\mathsf{K}_{D}$, the decomposition also implies that dissipation increases when $\mathsf{K}_{D}$ becomes nonnormal with its eigenvalues fixed.
This increase in dissipation due to the nonnormality of $\mathsf{K}_{D}$ was first proposed in Ref.~\cite{fyodorov2025nonorthogonal} (see SM~\cite{SuppMat}). However, the analytical proof of the increase has been limited to the asymptotic behavior of a specific model for  $N\to\infty$~\cite{fyodorov2025nonorthogonal} and two-dimensional models~\cite{sornette2025life,troude2026pseudo}. 
Our decomposition confirms the increase without relying on any specific models.

Equation~\eqref{oscillatory EPR} further decomposes $\sigma^{\mathrm{osc}}$ into contributions from each eigenvalue.
This decomposition of $\sigma^{\mathrm{osc}}$ may be related to another decomposition in Ref.~\cite{sekizawa2024decomposing}, which decomposes the EPR into the contributions of each eigenmode of $\tilde{\mathsf{K}}_{-}$.
This mode decomposition is also based on oscillatory behavior, but it does not explicitly address the impact of nonnormality. These two decompositions coincide when $\tilde{\mathsf{K}}$ is normal (see SM for details~\cite{SuppMat}).

\textit{Application 1: Stricter dissipation-coherence trade-off}.---%
We now demonstrate an application of our decomposition by deriving the DCT for OU processes with noise-induced oscillations.

The coherence of noisy oscillations is characterized by the long-time behavior of the autocorrelation function. For simplicity, we assume a unique slowest eigenmode, meaning that $\mathrm{Re}(\lambda_1)<\mathrm{Re}(\lambda_n)$ for any $\lambda_n\notin\{\lambda_1,\lambda_1^{\ast}\}$, where $\ast$ denotes the complex conjugate. This assumption and Eq.~\eqref{correlation spectral decomp.} in Appendix~C determine the long-time behavior of the autocorrelation function~\footnote{Even if $\tilde{\mathsf{K}}$ is not diagonalizable, we can obtain the same result using the Jordan normal form of $\tilde{\mathsf{K}}$.} as $C^a_t\sim\mathrm{e}^{-t/\tau_{\mathrm{c}}}\cos(\omega t+\theta_0)$, with $\tau_{\mathrm{c}}=\mathrm{Re}(\lambda_1)^{-1}$, $\omega\coloneqq|\mathrm{Im}(\lambda_1)|$, and an initial phase $\theta_0$. Here the symbol $\sim$ indicates asymptotic scaling up to sub-exponential prefactors.
We refer to $\tau_{\mathrm{c}}$, the time scale of exponential decay, as the correlation time. Using these quantities, we can define a measure of the coherence as
\begin{align}
    \mathcal{N}\coloneqq\tau_{\mathrm{c}}\times\frac{\omega}{2\pi}=\frac{|\mathrm{Im}(\lambda_1)|}{2\pi\mathrm{Re}(\lambda_1)},
    \label{Number of coherent oscillations}
\end{align}
which is the number of coherent oscillations~\cite{oberreiter2022universal,gaspard2002trace,morelli2007,dEysmond2013}.

We derive the DCT using the oscillatory-nonnormal decomposition. The decomposition immediately leads to 
\begin{align}
    \sigma^{\mathrm{st}}\geq\sum_{n=1}^N\frac{\mathrm{Im}(\lambda_n)^2}{\mathrm{Re}(\lambda_n)}\geq\frac{2\mathrm{Im}(\lambda_1)^2}{\mathrm{Re}(\lambda_1)}=2\omega^2\tau_{\mathrm{c}},
    \label{oscillatory bound}
\end{align}
where the second inequality is obtained by retaining the terms corresponding to $\lambda_1$ and $\lambda_1^{\ast}$ in the sum.
Using Eq.~\eqref{Number of coherent oscillations} and the oscillatory period $\tau_{\mathrm{p}}\coloneqq2\pi/\omega$, we obtain the DCT as
\begin{align}
    \tau_{\mathrm{p}}\sigma^{\mathrm{st}}\geq8\pi^2\mathcal{N}.
    \label{DCT}
\end{align}
Here, $\tau_{\mathrm{p}}\sigma^{\mathrm{st}}$ is the EP required for one oscillatory period.
This derivation implies that the equality in the DCT~\eqref{DCT} is achieved if and only if $\tilde{\mathsf{K}}$ is normal and $\mathrm{Im}(\lambda_n)$ vanishes for all eigenvalues other than $\lambda_1$ and $\lambda_1^{\ast}$.

We can regard the DCT in Eq.~\eqref{DCT} as a constraint on the leading eigenmode of $\mathcal{L}$. Indeed, $-\lambda_1$ and $-\lambda_1^{\ast}$ are the nonzero eigenvalues of $\mathcal{L}$ with the largest real part. This  implies that the long-time behavior of the autocorrelation function is determined by $-\lambda_1$ even for nonlinear observables.

The DCT for the OU process in Eq.~\eqref{DCT} is stricter than the DCTs conjectured for Markov jump processes~\cite{oberreiter2022universal} and proven for stochastic limit cycles in the weak-noise limit~\cite{santolin2025dissipation,nagayama2025duality,kolchinsky2025comment}. Indeed, the DCTs for those systems have $4\pi^{2}\mathcal{N}$ as the attainable lower bound for the EP~\footnote{For the MJPs, it is also conjectured that the DCT can be violated if $2\pi\mathcal{N}$ is less than $1$~\cite{oberreiter2022universal}.}, which is half as large as $8\pi^{2}\mathcal{N}$ in Eq.~\eqref{DCT}.
This fact implies that the minimum dissipation required to achieve a given $\mathcal{N}$ is twice as large in the OU process as in the other systems. 
This thermodynamic inefficiency may be attributed to the fact that the oscillations in the OU process are induced by noise rather than nonlinear dynamics.

\textit{Application 2: Relaxation speedup by nonnormality}.---%
It is known that nonequilibrium driving that preserves the target steady-state distribution can reduce the correlation time~\cite{hwang1993accelerating,lelievre2013optimal,Anton2022}. Here, we show that this reduction implies an unavoidable thermodynamic cost due to nonnormality. We do this by establishing a trade-off between $\sigma^{\mathrm{nn}}$ and the degree of reduction.

We consider the reduction of $\tau_{\mathrm{c}}$ relative to a \textit{reference} system, which is obtained by replacing $\mathsf{K}$ in Eq.~\eqref{Fokker--Planck eq} with
\begin{align}
    \mathsf{K}_{\mathrm{eq}}\coloneqq\frac{\mathsf{K}+\mathsf{V}\mathsf{K}^{\top}\mathsf{V}^{-1}}{2}.
    \label{equilibrium drift matrix}
\end{align}
The Lyapunov equation and Eq.~\eqref{equilibrium drift matrix} imply $\mathsf{K}_{\mathrm{eq}}\mathsf{V}+\mathsf{V}\mathsf{K}_{\mathrm{eq}}^{\top}-2\mathsf{D}=\mathsf{0}$. 
In the whitened coordinate system, the drift matrix of the reference system is symmetric as $\tilde{\mathsf{K}}_{\mathrm{eq}}\coloneqq\mathsf{V}^{-1/2}\mathsf{K}_{\mathrm{eq}}\mathsf{V}^{1/2}=\tilde{\mathsf{K}}_{+}$~\footnote{The last equality follows from $\mathsf{V}^{-1/2}\mathsf{K}_{\mathrm{eq}}\mathsf{V}^{1/2}=\{\mathsf{V}^{-1/2}\mathsf{K}\mathsf{V}^{1/2}+\mathsf{V}^{1/2}\mathsf{K}^{\top}\mathsf{V}^{-1/2}\}/2=(\tilde{\mathsf{K}}+\tilde{\mathsf{K}}^{\top})/2=\tilde{\mathsf{K}}_{+}$.}.
These facts imply that the reference system has the steady state of the original system as its equilibrium state. Thus, we can investigate the nonequilibrium effects in the original system by comparing it with the reference system. This construction of the reference system is the Fokker–Planck/OU analogue~\cite{qian2013decomposition} of additive reversibilization for Markov jump processes~\cite{fill1991eigenvalue, bremaud2013markov, sakai2016eigenvalue, kolchinsky2024thermodynamic} (see SM~\cite{SuppMat}).

We introduce the correlation time of the reference system. Let $\{\lambda_n^{\mathrm{eq}}\}_{n=1}^N$ denote the eigenvalues of $\mathsf{K}_{\mathrm{eq}}$. These eigenvalues are real and positive, since $\mathsf{K}_{\mathrm{eq}}$ is similar to the positive-definite matrix $\tilde{\mathsf{K}}_{+}$. Here, the eigenvalues are labeled so that $\lambda_1^{\mathrm{eq}}\leq\lambda_2^{\mathrm{eq}}\leq\cdots\leq\lambda_N^{\mathrm{eq}}$. As in the original system, the correlation time of the reference system is given by $\tau^{\mathrm{eq}}_{\mathrm{c}}\coloneqq1/\lambda_1^{\mathrm{eq}}$. With some linear algebra, we derive the inequality $\mathrm{Re}(\lambda_1)\geq\lambda_1^{\mathrm{eq}}$ (see Appendix~E), which implies
\begin{align}
    \tau_{\mathrm{c}}\leq\tau^{\mathrm{eq}}_{\mathrm{c}}.
    \label{correlation time reduction}
\end{align}
In this sense, nonequilibrium driving can reduce the correlation time.
Note that, as shown in SM~\cite{SuppMat}, $(\tau^{\mathrm{eq}}_{\mathrm{c}})^{-1}$ can also be understood as the decay rate of autocorrelation of the original system at short-time scales. Thus, we can also interpret Eq.~\eqref{correlation time reduction} as the difference between the decay rates of the autocorrelation in long-time and short-time regimes.

Let $\lambda_{\mathrm{max}}(\tilde{\mathsf{D}})$ denote the largest eigenvalue of $\tilde{\mathsf{D}}=\mathsf{V}^{-1/2}\mathsf{D}\mathsf{V}^{-1/2}$, which is the intensity of diffusion in the whitened coordinate system.
This quantity enables us to bound the nonnormal EPR with the difference between $\{\mathrm{Re}(\lambda_n)\}_{n=1}^{N}$ and $\{\lambda_n^{\mathrm{eq}}\}_{n=1}^{N}$ as
\begin{align}
    \sigma^{\mathrm{nn}}\geq\frac{1}{\lambda_{\mathrm{max}}(\tilde{\mathsf{D}})}\sum_{n=1}^N(\mathrm{Re}(\lambda_n)-\lambda_n^{\mathrm{eq}})^2,
    \label{bound on nonnormal EPR}
\end{align}
which is derived in Appendix~E. 
We also obtain
\begin{align}
    \sigma^{\mathrm{nn}}\geq\frac{N}{N-1}\frac{[\tau_{\mathrm{c}}^{-1}-(\tau^{\mathrm{eq}}_{\mathrm{c}})^{-1}]^2}{\lambda_{\mathrm{max}}(\tilde{\mathsf{D}})},
    \label{bound on correlation time}
\end{align}
by relaxing Eq.~\eqref{bound on nonnormal EPR} and using the definitions of the correlation times (see Appendix~E).
Combining Eq.~\eqref{bound on correlation time} with Eq.~\eqref{correlation time reduction} shows that the reduction of the correlation time $\tau_{\mathrm{c}}<\tau^{\mathrm{eq}}_{\mathrm{c}}$ requires positive $\sigma^{\mathrm{nn}}$ and thus the system's nonnormality.

This result may also be useful from an engineering perspective.
In thermodynamic computing~\cite{conte2019thermodynamic}, sampling data from the steady state of a linear Langevin system is essential to implement matrix operations~\cite{aifer2024thermodynamic,melanson2025thermodynamic}. Shortening the correlation time speeds up the computation by decreasing both the initial relaxation time to the steady state and the time required for the data samples to become uncorrelated. The decomposition [Eq.~\eqref{oscillatory-nonnormal decomposition}] and the bound on $\sigma^{\mathrm{nn}}$ suggest that, for a prescribed relaxation speedup, dissipation is minimized by avoiding oscillatory contributions, i.e., by using nonnormal $\tilde{\mathsf{K}}$ with real eigenvalues.

\begin{figure}
    \centering
    \includegraphics[width=\linewidth]{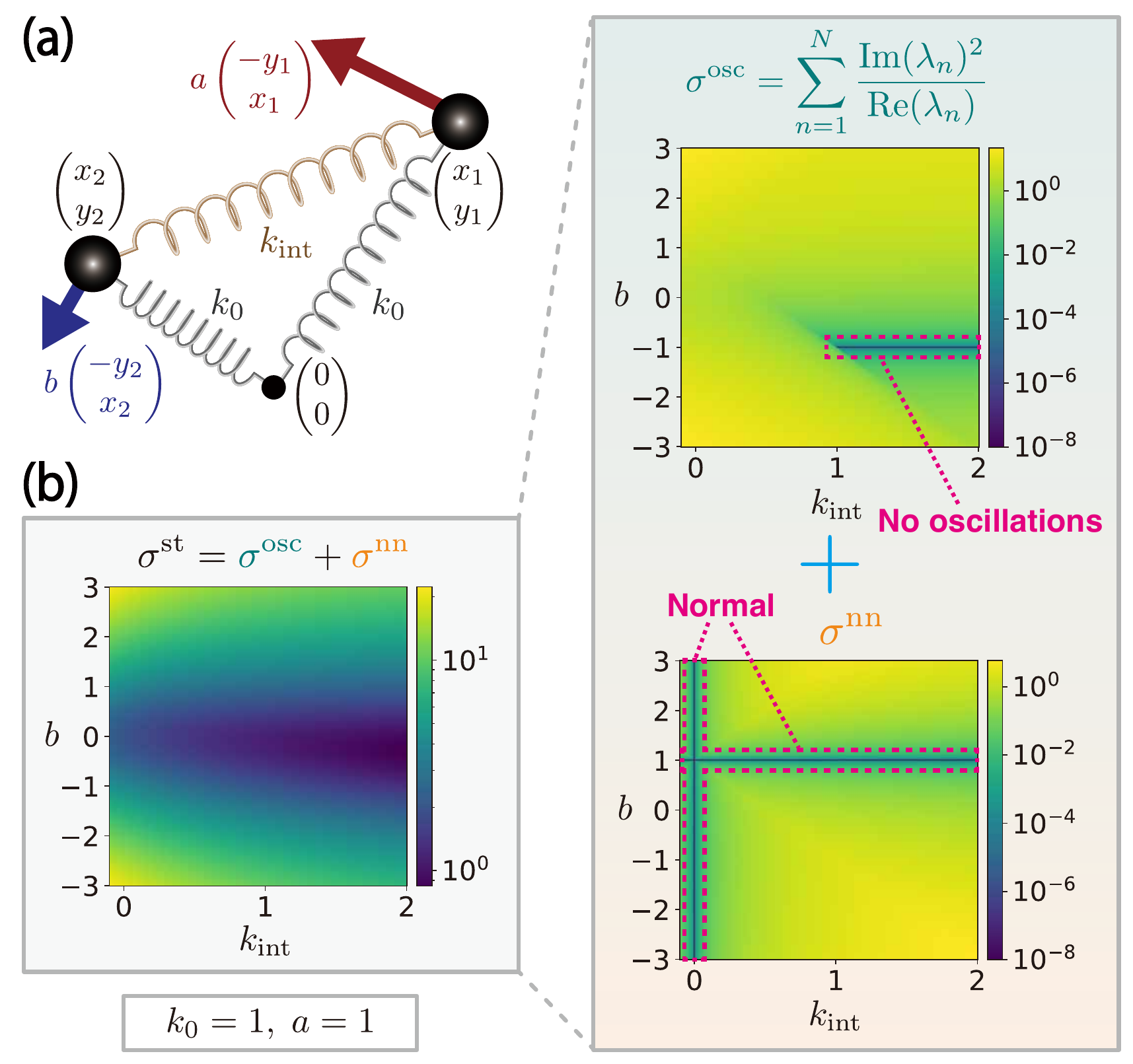}
    \caption{Numerical demonstration of the oscillatory-nonnormal decomposition. (a) The system used in the numerical demonstration. The parameters are fixed at $k_0 = 1$ and $a = 1$, while $k_{\mathrm{int}}$ and $b$ are varied. (b) $(k_{\mathrm{int}},b)$-dependence of the EPRs. The heatmaps of the EPRs use a logarithmic color scale with a finite lower cutoff.}
    \label{fig:numerics}
\end{figure}

\textit{Example}.---%
We demonstrate the oscillatory-nonnormal decomposition using two Brownian particles on a two-dimensional plane [Fig.~\ref{fig:numerics}(a)]. 
The position of each particle is denoted by $(x_i, y_i)^{\top}$ ($i=1,\,2$), and the state of the system is expressed as $\bm{x}=(x_1, y_1, x_2, y_2)^{\top}$. The two particles receive rotational forces $(-ay_1, ax_1)^{\top}$ and $(-by_2, bx_2)^{\top}$, respectively.
Each particle is connected to the origin of the plane by a spring with a spring constant $k_{0}$, and the two particles are connected to each other by a spring with a spring constant $k_{\mathrm{int}}$. All springs have a rest length of zero. 
For simplicity, we set the mobility and temperature of the medium to unity. Then, the time evolution of this system is governed by Eq.~\eqref{Fokker--Planck eq} with 
\begin{align}
    \mathsf{K}=\begin{pmatrix}
        k_0+k_{\mathrm{int}} & a & -k_{\mathrm{int}} & 0\\
        -a & k_0+k_{\mathrm{int}} & 0 & -k_{\mathrm{int}}\\
        -k_{\mathrm{int}} & 0 & k_0+k_{\mathrm{int}} & b\\
        0 & -k_{\mathrm{int}} & -b & k_0+k_{\mathrm{int}}
    \end{pmatrix},
\end{align}
and $\mathsf{D}=\mathsf{I}$~\cite{sekizawa2024decomposing}.
In the following, we set the parameters in $\mathsf{K}$ to $k_{0}=1$, $k_{\mathrm{int}}\in[-0.1,2]$, $a=1$, and $b\in[-3,3]$, where negative $k_{\mathrm{int}}$ implies repulsive coupling. We show the $(k_{\mathrm{int}},b)$-dependence of the EPRs in Fig.~\ref{fig:numerics}(b).

Let us focus on when $\sigma^{\mathrm{osc}}$ vanishes. As shown in Fig.~\ref{fig:numerics}(b), the oscillatory EPR vanishes when $b=-a=-1$ and $k_{\mathrm{int}}\ge 1$. This is verified by computing the eigenvalues of $\mathsf{K}$. The four eigenvalues are expressed with $(s_1, s_2) \in \{+1, -1\}^2$ as
\begin{align}
    \lambda_{s_1,s_2}=k_{0}+k_{\mathrm{int}}+s_1\sqrt{k_{\mathrm{int}}^2-\frac{(a-b)^2}{4}}+\mathrm{i}s_2\frac{a+b}{2}.
    \label{eigenvalues of the spring model}
\end{align}
When $b=-a$, $\lambda_{s_1,s_2}$ is complex-valued only if $k_{\mathrm{int}}^2 < (a-b)^2/4$, so that the square root becomes imaginary. Thus, if the coupling is sufficiently strong ($|k_{\mathrm{int}}| \geq |a-b|/2$) and $b=-a$, all eigenvalues become real and $\sigma^{\mathrm{osc}}$ vanishes.
This behavior corresponds to the cancellation of oppositely directed oscillations due to the coupling.

We also focus on the vanishing of $\sigma^{\mathrm{nn}}$, which occurs with either $k_{\mathrm{int}}=0$ or $b=a=1$. The former implies that the two oscillators are not coupled. The latter corresponds to the case where the angular frequencies of the two oscillators are the same.
These align with the intuition that nonnormality is induced by the coupling of different oscillatory modes.

\textit{Discussion}.---%
In this Letter, we derived the oscillatory-nonnormal decomposition of the steady-state EPR and applied it to obtain thermodynamic bounds for OU processes. An important challenge is to extend our results beyond linear systems. 
For example, the classification of steady states into four types might be extended to general Markov processes based on the eigenvalues and the nonnormality of the generator.
The reduction of correlation time by nonequilibrium driving has been studied in the context of stochastic thermodynamics~\cite{kolchinsky2024thermodynamic} and Markov chain Monte Carlo methods~\cite{ichiki2013violation,diaconis2000analysis,suwa2010markov,turitsyn2011irreversible,chen2013accelerating,sakai2016eigenvalue,bierkens2016non,takahashi2016conflict,kaiser2017acceleration,ghimenti2022accelerating}.  Based on these results, it may be possible to derive an inequality like Eq.~\eqref{bound on nonnormal EPR} and demonstrate that nonnormality is essential for reducing the correlation time even in general Markov processes.

\indent

\begin{acknowledgments}
\textit{Acknowledgments}.---%
The authors thank Naruo Ohga for discussions.
R.N. thanks Guo-Hua Xu and Ruicheng Bao for helpful comments.
Several proofs were derived with assistance from Gemini and ChatGPT.
R.N.\ is supported by JSPS KAKENHI Grants No.~25KJ0931 and JSR Fellowship, the University of Tokyo.
A.K.\ is partly supported by the John Templeton Foundation (grant 62828) and by the European Union's Horizon 2020 research and innovation programme under the Marie Sk{\l}odowska-Curie Grant Agreement No.~101068029. 
S.I.\ is supported by JSPS KAKENHI Grants No.~23K22412, No.~23H00467, and No.~25K24775, 
JST ERATO Grant No.~JPMJER2302, 
and UTEC-UTokyo FSI Research Grant Program.
\end{acknowledgments}

\input{biblio.bbl}

\newpage

\onecolumngrid

\newpage

\section*{End Matter}

\twocolumngrid

\textit{Appendix A: Derivation of Eq.~\eqref{EPR simple form}}.---%
Here, we derive Eq.~\eqref{EPR simple form}. Using the stationary distribution $p^{\mathrm{st}}(\bm{x})=[(2\pi)^N\det(\mathsf{V})]^{-1/2}\exp{(-\bm{x}^{\top}\mathsf{V}^{-1}\bm{x}/2)}$, the flux field is given by $\bm{j}^{\mathrm{st}}(\bm{x})=-p^{\mathrm{st}}(\bm{x})(\mathsf{K}-\mathsf{D}\mathsf{V}^{-1})\bm{x}$. This expression rewrites the definition of $\sigma^{\mathrm{st}}$~\cite{landi2013entropy,godreche2018characterising} as
\begin{align}
    \sigma^{\mathrm{st}}&=\int d\bm{x}\,p^{\mathrm{st}}(\bm{x})\bm{x}^{\top}(\mathsf{K}-\mathsf{D}\mathsf{V}^{-1})^{\top}\mathsf{D}^{-1}(\mathsf{K}-\mathsf{D}\mathsf{V}^{-1})\bm{x}\notag\\
    &=\sum_{\alpha,\beta}V_{\beta\alpha}[(\mathsf{K}-\mathsf{D}\mathsf{V}^{-1})^{\top}\mathsf{D}^{-1}(\mathsf{K}-\mathsf{D}\mathsf{V}^{-1})]_{\alpha\beta}\notag\\
    &=\mathrm{tr}[\mathsf{V}(\mathsf{K}-\mathsf{D}\mathsf{V}^{-1})^{\top}\mathsf{D}^{-1}(\mathsf{K}-\mathsf{D}\mathsf{V}^{-1})],
    \label{EPR trace form}
\end{align}
where we used $\int d\bm{x}\,x_{\alpha}x_{\beta}p^{\mathrm{st}}(\bm{x})=V_{\alpha\beta}=V_{\beta\alpha}$ and $V_{\alpha\beta}$ denotes the $(\alpha,\beta)$th component of $\mathsf{V}$.
Using $\tilde{\mathsf{K}}_{+}=\mathsf{V}^{-1/2}\mathsf{D}\mathsf{V}^{-1/2}$ and $\tilde{\mathsf{K}}_{-}=\mathsf{V}^{-1/2}(\mathsf{K}\mathsf{V}-\mathsf{D})\mathsf{V}^{-1/2}$ in Eq.~\eqref{EPR trace form}, we immediately obtain Eq.~\eqref{EPR simple form}.

\textit{Appendix B: Derivation of the equivalence of normality}.---%
We first show that $\tilde{\mathsf{K}}$ is normal if and only if $[\mathsf{K},\mathsf{D}\mathsf{V}^{-1}]=\mathsf{0}$. Since $[\tilde{\mathsf{K}}_{+},\tilde{\mathsf{K}}]=([\tilde{\mathsf{K}}+\tilde{\mathsf{K}}^{\top},\tilde{\mathsf{K}}])/2=[\tilde{\mathsf{K}}^{\top},\tilde{\mathsf{K}}]/2$, the normality of $\tilde{\mathsf{K}}$ is equivalent to $[\tilde{\mathsf{K}}_{+},\tilde{\mathsf{K}}]=\mathsf{0}$.
Using $\tilde{\mathsf{K}}=\mathsf{V}^{-1/2}\mathsf{K}\mathsf{V}^{1/2}$ and $\tilde{\mathsf{K}}_{+}=\mathsf{V}^{-1/2}\mathsf{D}\mathsf{V}^{-1/2}$, we obtain $[\mathsf{K},\mathsf{D}\mathsf{V}^{-1}]=[\mathsf{V}^{1/2}\tilde{\mathsf{K}}\mathsf{V}^{-1/2},\mathsf{V}^{1/2}\tilde{\mathsf{K}}_{+}\mathsf{V}^{-1/2}]=\mathsf{V}^{1/2}[\tilde{\mathsf{K}}, \tilde{\mathsf{K}}_{+}]\mathsf{V}^{-1/2}$. This relation and the positive definiteness of $\mathsf{V}$ ensure the equivalence of $[\tilde{\mathsf{K}}_{+},\tilde{\mathsf{K}}]=\mathsf{0}$ and $[\mathsf{K},\mathsf{D}\mathsf{V}^{-1}]=\mathsf{0}$. Thus, the normality of $\tilde{\mathsf{K}}$ is equivalent to $[\mathsf{K},\mathsf{D}\mathsf{V}^{-1}]=\mathsf{0}$.

Next, we show that $\mathsf{K}_D=\mathsf{D}^{-1/2}\mathsf{K}\mathsf{D}^{1/2}$ is normal if and only if $[\mathsf{K},\mathsf{V}\mathsf{D}^{-1}]=\mathsf{0}$. 
We use the covariance matrix in the coordinate system where the diffusion matrix becomes the identity. 
This covariance matrix is given by $\mathsf{V}_D\coloneqq\mathsf{D}^{-1/2}\mathsf{V}\mathsf{D}^{-1/2}$ and rewrites the Lyapunov equation as $\mathsf{K}_D\mathsf{V}_D+\mathsf{V}_D\mathsf{K}_D^{\top}-2\mathsf{I}=\mathsf{0}$. 
The normality of $\mathsf{K}_D$ is equivalent to $[\mathsf{K}_D,\mathsf{V}_D]=\mathsf{0}$ as shown below.
Assume that $\mathsf{K}_D$ is normal, i.e., $[\mathsf{K}_D,\mathsf{K}_D^{\top}]=\mathsf{0}$. 
Then, $[\mathsf{K}_D,\mathsf{V}_D]=\mathsf{0}$ holds because $\mathsf{V}_D$ is rewritten as $2\int_{0}^{\infty}dt\,\mathrm{e}^{-\mathsf{K}_Dt}\mathrm{e}^{-\mathsf{K}_D^{\top}t}$~\cite{risken1989fokker}, and the assumption makes $\mathsf{K}_D$ commute with both $\mathrm{e}^{-\mathsf{K}_Dt}$ and $\mathrm{e}^{-\mathsf{K}_D^{\top}t}$.
Conversely, assume that $\mathsf{K}_D$ commutes with $\mathsf{V}_D$. This assumption reduces the Lyapunov equation to $\mathsf{V}_D(\mathsf{K}_D+\mathsf{K}_D^{\top})=2\mathsf{I}$, which implies $\mathsf{K}_D+\mathsf{K}_D^{\top}=2\mathsf{V}_D^{-1}$. This relation leads to $[\mathsf{K}_D,\mathsf{K}_D^{\top}]=-[\mathsf{K}_D,\mathsf{K}_D-2\mathsf{V}_D^{-1}]=2[\mathsf{K}_D,\mathsf{V}_D^{-1}]$, which vanishes due to the assumption.
Using the definitions of $\mathsf{K}_D$ and $\mathsf{V}_D$, we obtain $[\mathsf{K}_D,\mathsf{V}_D]=[\mathsf{D}^{-1/2}\mathsf{K}\mathsf{D}^{1/2},\mathsf{D}^{-1/2}\mathsf{V}\mathsf{D}^{-1/2}]=\mathsf{D}^{-1/2}[\mathsf{K},\mathsf{V}\mathsf{D}^{-1}]\mathsf{D}^{1/2}$. The positive definiteness of $\mathsf{D}$ ensures that $[\mathsf{K},\mathsf{V}\mathsf{D}^{-1}]=\mathsf{0}$ is equivalent to $[\mathsf{K}_D,\mathsf{V}_D]=\mathsf{0}$, and thus equivalent to the normality of $\mathsf{K}_D$.

Since $\mathsf{D}\mathsf{V}^{-1}$ is the inverse of $\mathsf{V}\mathsf{D}^{-1}$, $[\mathsf{K},\mathsf{D}\mathsf{V}^{-1}]=\mathsf{0}$ and $[\mathsf{K},\mathsf{V}\mathsf{D}^{-1}]=\mathsf{0}$ are equivalent. Thus, the normality of $\tilde{\mathsf{K}}$ is equivalent to the normality of $\mathsf{K}_D$.


\textit{Appendix C: Details of the autocorrelation function}.---%
We first define the correlation matrix in the whitened coordinate system as $\mathsf{C}_t\coloneqq\langle\tilde{\bm{x}}_t\tilde{\bm{x}}_0^{\top}\rangle_{\mathrm{st}}$~\cite{kubo1966fluctuation}. To calculate this matrix, we consider the Langevin equation corresponding to Eq.~\eqref{Fokker--Planck eq}, $\dot{\bm{x}}_t=-\mathsf{K}\bm{x}_t+\sqrt{2}\mathsf{G}\bm{\xi}_t$,
where $\bm{x}_t$ is the state of the system at time $t$, and $\mathsf{G}$ satisfies $\mathsf{G}\mathsf{G}^{\top}=\mathsf{D}$. The vector $\bm{\xi}_t$ is white Gaussian noise, where $\langle\bm{\xi}_t\rangle=\bm{0}$ and $\langle\bm{\xi}_t\bm{\xi}_{t'}^{\top}\rangle=\delta(t-t')\mathsf{I}$. Here, $\langle\cdots\rangle$ stands for the expected value and $\bm{0}$ denotes the zero vector.
In the whitened coordinate system, the state $\tilde{\bm{x}}_t=\mathsf{V}^{-1/2}\bm{x}_t$ evolves according to $\dot{\tilde{\bm{x}}}_t=-\tilde{\mathsf{K}}\tilde{\bm{x}}_t+\sqrt{2}\tilde{\mathsf{G}}\bm{\xi}_t$ with $\tilde{\mathsf{G}}\coloneqq\mathsf{V}^{-1/2}\mathsf{G}$. 
Using this Langevin equation, we obtain $\dot{\mathsf{C}}_t=\langle \dot{\tilde{\bm{x}}}_t\tilde{\bm{x}}_0^{\top}\rangle_{\mathrm{st}}=\langle (-\tilde{\mathsf{K}}\tilde{\bm{x}}_t+\sqrt{2}\tilde{\mathsf{G}}\bm{\xi}_t)\tilde{\bm{x}}_0^{\top}\rangle_{\mathrm{st}}=-\tilde{\mathsf{K}}\mathsf{C}_t$ for $t\geq0$. Here, we used the fact that $\bm{\xi}_t$ for $t\geq0$ is zero-mean and independent of $\tilde{\bm{x}}_0$. $\mathsf{C}_0=\langle \tilde{\bm{x}}_0\tilde{\bm{x}}_0^{\top}\rangle_{\mathrm{st}}$ is the steady-state covariance matrix in the whitened coordinate, so $\mathsf{C}_0=\tilde{\mathsf{V}}=\int d\bm{x}\,p^{\mathrm{st}}(\bm
{x})\mathsf{V}^{-1/2}\bm{x}\bm{x}^{\top}\mathsf{V}^{-1/2}=\mathsf{V}^{-1/2}\mathsf{V}\mathsf{V}^{-1/2}=\mathsf{I}$. Thus, the linear time evolution of the correlation matrix is solved by $\mathsf{C}_t=\mathrm{e}^{-\tilde{\mathsf{K}}t}$.

Using this correlation matrix, we can relate the autocorrelation function of $a(\tilde{\bm{x}})=\bm{a}^{\top}\tilde{\bm{x}}$ to $\tilde{\mathsf{K}}$ as 
\begin{align}
    C_t^a=\bm{a}^{\top}\mathsf{C}_t\bm{a}=\bm{a}^{\top}\mathrm{e}^{-\tilde{\mathsf{K}}t}\bm{a}.
    \label{correlation relation}
\end{align}
Based on this expression of $C_t^a$, we show the effect of the complex eigenvalues.
Let us temporarily assume that $\tilde{\mathsf{K}}$ is diagonalizable as $\tilde{\mathsf{K}}=\sum_{n=1}^N\lambda_n\bm{u}_n\bm{v}_n^{\top}$. Here, $\bm{v}_n$ and $\bm{u}_n$ represent the left and right eigenvectors of $\tilde{\mathsf{K}}$ corresponding to $\lambda_n$, which satisfy $\bm{v}_n^{\top}\bm{u}_m=\delta_{nm}$.
This spectral decomposition rewrites $C_t^a$ as 
\begin{align}
    C^a_t=\sum_{n=1}^{N}\mathrm{e}^{-\mathrm{Re}(\lambda_n)t-\mathrm{i}\mathrm{Im}(\lambda_n)t}(\bm{a}^{\top}\bm{u}_n\bm{v}_n^{\top}\bm{a}),
    \label{correlation spectral decomp.}
\end{align}
which is a superposition of modes that oscillate at a frequency $|\mathrm{Im}(\lambda_n)|/(2\pi)$ while exponentially decaying with a time constant $1/\mathrm{Re}(\lambda_n)$ [Fig.~\ref{fig:schematics}(a)]. Thus, complex eigenvalues induce oscillatory behavior in $C_t^a$.

Next, we show the effect of the nonnormality. With nonnormal $\tilde{\mathsf{K}}$, the initial decay of $C_t^a$ may become slower than the long-time decay~\cite{trefethen2020spectra}. Here, we define the \emph{decay rate} of $C_t^a$ at a timescale $\tau$ as $\gamma^{a}_{\tau}\coloneqq\liminf_{t\to\tau}t^{-1}\ln|{C^{a}_{0}}/{C^{a}_{t}}|$ so that $C^{a}_{t}$ decays as $e^{-\gamma^{a}_{\tau}t}$ when $t\simeq\tau$. In particular, the short-time decay rate $\gamma^{a}_{0}$ is obtained as $\gamma^a_0=\bm{a}^{\top}\tilde{\mathsf{K}}_{+}\bm{a}/(\bm{a}^{\top}\bm{a})$ using the Taylor expansion of Eq.~\eqref{correlation relation} (see SM~\cite{SuppMat}). On the other hand, Eq.~\eqref{correlation spectral decomp.} provides the long-time decay rate as $\gamma_{\infty}^a=\mathrm{Re}(\lambda_1)$. Here and in the following, we assume that $\bm{a}$ has nonzero overlap with the slowest eigenmode.

If $\tilde{\mathsf{K}}$ is normal, the eigenvalues of $\tilde{\mathsf{K}}_{+}$ are given by $\{\mathrm{Re}(\lambda_n)\}_{n=1}^N$, in which $\mathrm{Re}(\lambda_1)$ is the smallest. Since $\tilde{\mathsf{K}}_{+}$ is symmetric, the Courant variational principle~\cite{Courant1989} leads to $\gamma^a_0=\bm{a}^{\top}\tilde{\mathsf{K}}_{+}\bm{a}/(\bm{a}^{\top}\bm{a})\geq\mathrm{Re}(\lambda_1)=\gamma^a_{\infty}$. This inequality implies that, in normal systems, the initial decay of $C_t^a$ is faster than (or equal to) its asymptotic decay. If $\tilde{\mathsf{K}}$ is nonnormal, this inequality is not valid in general, and the initial decay can be slower than the decay in the long-time regime. Indeed, taking $\bm{a}$ as the eigenvector of $\tilde{\mathsf{K}}_{+}$ corresponding to its smallest eigenvalue $\lambda_{\mathrm{min}}(\tilde{\mathsf{K}}_{+})$, we obtain $\gamma_0^a=\lambda_{\mathrm{min}}(\tilde{\mathsf{K}}_{+})\leq\mathrm{Re}(\lambda_1)=\gamma^a_{\infty}$ as follows: We introduce the normalized eigenvector $\hat{\bm{u}}_1\coloneqq\bm{u}_1/\|\bm{u}_1\|$. 
Due to $\hat{\bm{u}}_1^{\dag}\hat{\bm{u}}_1=1$ and $\tilde{\mathsf{K}}\hat{\bm{u}}_1=\lambda_1\hat{\bm{u}}_1$, the eigenvalue $\lambda_1$ is expressed as $\lambda_1=\hat{\bm{u}}_1^{\dag}\tilde{\mathsf{K}}\hat{\bm{u}}_1$. We may then write $\mathrm{Re}(\lambda_1)=\{\hat{\bm{u}}_1^{\dag}\tilde{\mathsf{K}}\hat{\bm{u}}_1+(\hat{\bm{u}}_1^{\dag}\tilde{\mathsf{K}}\hat{\bm{u}}_1)^{\ast}\}/2=\hat{\bm{u}}_1^{\dag}(\tilde{\mathsf{K}}+\tilde{\mathsf{K}}^{\dag})\hat{\bm{u}}_1/2=\hat{\bm{u}}_1^{\dag}\tilde{\mathsf{K}}_{+}\hat{\bm{u}}_1$, where $\tilde{\mathsf{K}}^{\dag}=\tilde{\mathsf{K}}^{\top}$ is also used.
The Courant variational principle bounds $\hat{\bm{u}}_1^{\dag}\tilde{\mathsf{K}}_{+}\hat{\bm{u}}_1$ with the smallest eigenvalue of $\tilde{\mathsf{K}}_{+}$ as $\mathrm{Re}(\lambda_1)=\hat{\bm{u}}_1^{\dag}\tilde{\mathsf{K}}_{+}\hat{\bm{u}}_1\geq\lambda_{\mathrm{min}}(\tilde{\mathsf{K}}_+)$.

Note that if $\tilde{\mathsf{K}}$ is not diagonalizable, we can no longer write $C_t^a$ as a sum of exponentially decaying eigenmodes as in Eq.~\eqref{correlation spectral decomp.}. Instead, the decomposition will include modes with polynomial prefactors, reflecting the presence of nontrivial Jordan blocks of $\tilde{\mathsf{K}}$.

\textit{Appendix D: Details of the decomposition}.---%
We first show that $\tilde{\mathsf{K}}$ is normal if and only if $\mathsf{S}\in\mathsf{H}\mathcal{D}$. Recall that $\tilde{\mathsf{K}}$ is normal if and only if $\mathsf{T}$ is diagonal, hence also $\mathsf{S}$ and $\mathsf{H}$ are diagonal. Since $\mathsf{H}$ is positive-definite, it is invertible. Then, $\mathsf{H}^{-1}\mathsf{S}$ is also diagonal, which implies $\mathsf{S}\in\mathsf{H}\mathcal{D}$.
To show the converse, we assume that  $\mathsf{S}\in\mathsf{H}\mathcal{D}$, meaning  there exists a diagonal matrix $\mathsf{X}$ that satisfies $\mathsf{S}=\mathsf{H}\mathsf{X}$. In component form, this relation is expressed as $S_{nm}=H_{nm}X_{mm}$. Focusing on the case of $n=m$, we can see that $X_{nn}$ is either imaginary or zero because of the relations ${S}_{nn}=\mathrm{i}\mathrm{Im}(\lambda_n)$ and ${H}_{nn}=\mathrm{Re}(\lambda_n)>0$.
Since $\mathsf{H}+\mathsf{S}=\mathsf{T}$ is upper triangular, we obtain ${H}_{nm}+S_{nm}=(1+{X}_{mm}){H}_{nm}=0$ for $n>m$. Then, since $X_{mm}$ is either imaginary or zero, $1+{X}_{mm}\neq0$ and $H_{nm}=0$. Since $\mathsf H$ is Hermitian, we have  $H_{nm}=H_{mn}^{\ast}$ and therefore ${H}_{nm}=0$ for all $n\neq m$, which implies that $\mathsf{H}$ is diagonal. This implies that $\mathsf{T}$ is diagonal and $\tilde{\mathsf{K}}$ is normal.

Next, we derive the expression in Eq.~\eqref{oscillatory EPR}.
To begin, we consider the orthogonal complement of $\mathsf{H}\mathcal{D}$ with respect to the inner product $\langle\cdot,\cdot\rangle_{\mathsf{H}^{-1}}$. This orthogonal complement,  denoted $(\mathsf{H}\mathcal{D})^{\perp}$, is the space of \emph{hollow} matrices, i.e., matrices whose diagonal elements are all equal to $0$. Indeed, a matrix $\mathsf{Y}$ belongs to $(\mathsf{H}\mathcal{D})^{\perp}$ if and only if $0=\langle\mathsf{Y},\mathsf{H}\mathsf{X}\rangle_{\mathsf{H}^{-1}}=\sum_{n}{Y}^{\ast}_{nn}X_{nn}$ for any $\mathsf{X}\in\mathcal{D}$, which immediately implies ${Y}_{nn}=0$ for all $n$.

Using this property of $(\mathsf{H}\mathcal{D})^{\perp}$ and $\mathsf{S}-\mathsf{S}^{\star}\in(\mathsf{H}\mathcal{D})^{\perp}$, we obtain $S^{\star}_{nn}=S_{nn}=\mathrm{i}\mathrm{Im}(\lambda_n)$. On the other hand, $\mathsf{S}^{\star}\in\mathsf{H}\mathcal{D}$ also enables us to express $\mathsf{S}^{\star}$ as $\mathsf{H}\mathsf{X}^{\star}$ with a diagonal matrix $\mathsf{X}^{\star}$, which leads to ${S}^{\star}_{nn}={H}_{nn}X^{\star}_{nn}=\mathrm{Re}(\lambda_n){X}^{\star}_{nn}$. Combining these two expressions of ${S}^{\star}_{nn}$ implies $X^{\star}_{nn}=\mathrm{i}\mathrm{Im}(\lambda_n)/\mathrm{Re}(\lambda_n)$. Using this $\mathsf{X}^{\star}$, we obtain $\|\mathsf{S}^{\star}\|^2_{\mathsf{H}^{-1}}=\mathrm{tr}(\mathsf{X}^{\star\dag}\mathsf{H}^{\dag}\mathsf{H}^{-1}\mathsf{H}\mathsf{X}^{\star})=\mathrm{tr}(\mathsf{X}^{\star\dag}\mathsf{H}\mathsf{X}^{\star})=\sum_{n=1}^{N}\mathrm{Im}(\lambda_n)^2/\mathrm{Re}(\lambda_n)$, where we also used $\mathsf{H}=\mathsf{H}^{\dag}$.

\textit{Appendix E: Details of the bound on $\sigma^{\mathrm{nn}}$}.---%
First, we derive $\mathrm{Re}(\lambda_1)\geq\lambda_1^{\mathrm{eq}}$, which implies Eq.~\eqref{correlation time reduction}. Since $\tilde{\mathsf{K}}_{+}$ is similar to $\mathsf{K}_{\mathrm{eq}}$ as $\tilde{\mathsf{K}}_{+}=\mathsf{V}^{-1/2}\mathsf{K}_{\mathrm{eq}}\mathsf{V}^{1/2}$, we obtain $\lambda_{\mathrm{min}}(\tilde{\mathsf{K}}_{+})=\lambda_1^{\mathrm{eq}}$. Combining this relation and the inequality $\mathrm{Re}(\lambda_1)\geq\lambda_{\mathrm{min}}(\tilde{\mathsf{K}}_+)$ derived in Appendix~C implies the desired inequality.

Second, we derive the bound in Eq.~\eqref{bound on nonnormal EPR}.
We start with the expression $\sigma^{\mathrm{nn}}=\|\mathsf{S}-\mathsf{S}^{\star}\|^2_{\mathsf{H}^{-1}}$. Since $\mathsf{H}$ is similar to $\mathsf{K}_{\mathrm{eq}}$ as $\mathsf{H}=\mathsf{U}^{\dag}\tilde{\mathsf{K}}_{+}\mathsf{U}=\mathsf{U}^{\dag}\mathsf{V}^{-1/2}\mathsf{K}_{\mathrm{eq}}\mathsf{V}^{1/2}\mathsf{U}$, the eigenvalues of $\mathsf{H}^{-1}$ are given by $\{1/\lambda^{\mathrm{eq}}_n\}_{n=1}^{N}$, where $1/\lambda^{\mathrm{eq}}_N$ is the smallest. Using this fact and the positive definiteness of $\mathsf{H}^{-1}$, we obtain
\begin{align}
    \sigma^{\mathrm{nn}}\geq\frac{1}{\lambda^{\mathrm{eq}}_N}\|\mathsf{S}-\mathsf{S}^{\star}\|^2_{\mathrm{F}}=\frac{1}{\lambda_{\mathrm{max}}(\tilde{\mathsf{D}})}\|\mathsf{S}-\mathsf{S}^{\star}\|^2_{\mathrm{F}},
    \label{nn bound derivation 1}
\end{align}
with the Frobenius norm defined as $\|\mathsf{M}\|_{\mathrm{F}}\coloneqq\sqrt{\mathrm{tr}(\mathsf{M}^{\dag}\mathsf{M})}$. Here, $\lambda^{\mathrm{eq}}_N=\lambda_{\mathrm{max}}(\tilde{\mathsf{D}})$ follows from $\tilde{\mathsf{K}}_{+}=\mathsf{V}^{-1/2}\mathsf{D}\mathsf{V}^{-1/2}=\tilde{\mathsf{D}}$.
Recalling that $\mathsf{S}-\mathsf{S}^{\star}$ is a hollow matrix and $\mathsf{S}^{\star}=\mathsf{H}\mathsf{X}^{\star}$, we can calculate $\|\mathsf{S}-\mathsf{S}^{\star}\|^2_{\mathrm{F}}$ as
\begin{align}
    \|\mathsf{S}-\mathsf{S}^{\star}\|^2_{\mathrm{F}}&=\sum_{n}\sum_{m(\neq n)}|{S}_{nm}-S^{\star}_{nm}|^2\notag\\
    &=\sum_{n}\sum_{m(\neq n)}|{S}_{nm}-{H}_{nm}X^{\star}_{mm}|^2\notag\\
    &=\sum_{n}\sum_{m(\neq n)}|{H}_{nm}|^2|\pm{1}-{X}^{\star}_{mm}|^2.
\end{align}
In the last transformation, we also used the fact that $\mathsf{T}$ is upper triangular, i.e., ${T}_{nm}=0$ for $n>m$. Indeed, this fact leads to ${S}_{nm}={T}_{nm}/2={H}_{nm}$ for $n<m$ and ${S}_{nm}=-{T}_{mn}^{\ast}/2=-{H}_{nm}$ for $n>m$. Since ${X}^{\star}_{mm}=\mathrm{i}\mathrm{Im}(\lambda_m)/\mathrm{Re}(\lambda_m)$ is imaginary, we obtain $|\pm{1}-{X}^{\star}_{mm}|^2=1+|{X}^{\star}_{mm}|^2\geq1$. This implies $\|\mathsf{S}-\mathsf{S}^{\star}\|^2_{\mathrm{F}}\geq\sum_{n}\sum_{m(\neq n)}|{H}_{nm}|^2$. Defining $\mathsf{H}^{\mathrm{diag}}\in\mathcal{D}$ as ${H}^{\mathrm{diag}}_{nm}\coloneqq {H}_{nn}\delta_{nm}$, we can rewrite and relax this bound on $\|\mathsf{S}-\mathsf{S}^{\star}\|^2_{\mathrm{F}}$ as
\begin{align}
    \|\mathsf{S}-\mathsf{S}^{\star}\|^2_{\mathrm{F}}\geq\|\mathsf{H}^{\mathrm{diag}}-\mathsf{H}\|^2_{\mathrm{F}}\geq\sum_{n=1}^N(\mathrm{Re}(\lambda_n)-\lambda_n^{\mathrm{eq}})^2.
    \label{nn bound derivation 2}
\end{align}
Here, the last inequality is derived from the Hoffman--Wielandt theorem~\cite[Corollary~6.3.8]{Horn1985} as follows. Let $\{\lambda_n(\mathsf{H}')\}_{n=1}^N$ denote the eigenvalues of an $N\times N$ Hermitian matrix $\mathsf{H}'$ labeled so that $\lambda_1(\mathsf{H}')\leq\lambda_2(\mathsf{H}')\leq\cdots\leq\lambda_N(\mathsf{H}')$. Because $\mathsf{H}$ and $\mathsf{H}+(\mathsf{H}^{\mathrm{diag}}-\mathsf{H})=\mathsf{H}^{\mathrm{diag}}$ are Hermitian, this theorem implies $\|\mathsf{H}^{\mathrm{diag}}-\mathsf{H}\|^2_{\mathrm{F}}\geq \sum_{n=1}^N(\lambda_n(\mathsf{H}^{\mathrm{diag}})-\lambda_n(\mathsf{H}))^2$. The definition of $\mathsf{H}^{\mathrm{diag}}$ leads to $\lambda_n(\mathsf{H}^{\mathrm{diag}})=\mathrm{Re}(\lambda_n)$. The similarity between $\tilde{\mathsf{K}}_+$ and $\mathsf{H}$ also yields $\lambda_n(\mathsf{H})=\lambda_n^{\mathrm{eq}}$. Thus, equation~\eqref{nn bound derivation 2} is obtained. Combining this result with Eq.~\eqref{nn bound derivation 1}, we obtain Eq.~\eqref{bound on nonnormal EPR}.

Third, we derive Eq.~\eqref{bound on correlation time}. We assume $N\geq2$. Since $\mathrm{tr}(\tilde{\mathsf{K}})=\mathrm{tr}(\tilde{\mathsf{K}}_+)$, we have $\sum_{n=1}^N\lambda_n^{\mathrm{eq}}=\sum_{n=1}^N\lambda_n=\sum_{n=1}^N\mathrm{Re}(\lambda_n)$, where we also used $\sum_{n=1}^N\mathrm{Im}(\lambda_n)=0$. This relation and the Cauchy--Schwarz inequality lead to
\begin{align}
    \sum_{n=2}^N(\mathrm{Re}(\lambda_n)-\lambda_n^{\mathrm{eq}})^2&\geq\frac{\left[\sum_{n=2}^N(\mathrm{Re}(\lambda_n)-\lambda_n^{\mathrm{eq}})\right]^2}{\sum_{n=2}^N 1^2}\notag\\
    &=\frac{(\mathrm{Re}(\lambda_1)-\lambda_1^{\mathrm{eq}})^2}{N-1}.
\end{align}
Using this inequality in Eq.~\eqref{bound on nonnormal EPR}, we obtain Eq.~\eqref{bound on correlation time}.

\newpage

\pagebreak
\setcounter{secnumdepth}{2}
\onecolumngrid
\begin{center}
\textbf{\large Supplemental Material for\\``Oscillatory-nonnormal decomposition of dissipation in Ornstein--Uhlenbeck processes''}
\end{center}
\setcounter{section}{0}
\setcounter{equation}{0}
\setcounter{figure}{0}
\setcounter{table}{0}
\setcounter{page}{1}
\makeatletter
\renewcommand{\theequation}{S\arabic{equation}}
\renewcommand{\thefigure}{S\arabic{figure}}
\renewcommand{\thesection}{\Roman{section}}

\section{Equivalence between the normality of $\tilde{\mathsf{K}}$ and that of the twisted generator $\tilde{\mathcal{L}}$}
Here, we show that the normality of $\tilde{\mathsf{K}}$ is equivalent to that of $\tilde{\mathcal{L}}$, which is defined as
\begin{align}
    \tilde{\mathcal{L}}[\psi(\bm{x})]\coloneqq p^{\mathrm{st}}(\bm{x})^{-\frac{1}{2}}\mathcal{L}[p^{\mathrm{st}}(\bm{x})^{\frac{1}{2}}\psi(\bm{x})].
\end{align}

Using $p^{\mathrm{st}}(\bm{x})\propto\exp(-\bm{x}^{\top}\mathsf{V}^{-1}\bm{x}/2)$ and 
\begin{align}
    \mathcal{L}[p(\bm{x})]&=\bm{\nabla}\cdot(p(\bm{x})\mathsf{K}\bm{x})+\bm{\nabla}\cdot(\mathsf{D}\bm{\nabla}p(\bm{x}))\notag\\
    &=(\mathsf{K}\bm{x})\cdot\bm{\nabla}p(\bm{x})+\bm{\nabla}\cdot(\mathsf{D}\bm{\nabla}p(\bm{x}))+\mathrm{tr}(\mathsf{K})p(\bm{x}),
\end{align}
we can express $\tilde{\mathcal{L}}$ as
\begin{align}
    \tilde{\mathcal{L}}[\psi(\bm{x})]&=\mathcal{L}[\psi(\bm{x})]-(\mathsf{D}\mathsf{V}^{-1}\bm{x})\cdot\bm{\nabla}\psi(\bm{x})+\left\{\frac{\bm{x}^{\top}\mathsf{V}^{-1}\mathsf{D}\mathsf{V}^{-1}\bm{x}}{4}-\frac{\bm{x}^{\top}\mathsf{V}^{-1}\mathsf{K}\bm{x}}{2}-\frac{\mathrm{tr}(\mathsf{D}\mathsf{V}^{-1})}{2}\right\}\psi(\bm{x})\notag\\
    &=[(\mathsf{K}-\mathsf{D}\mathsf{V}^{-1})\bm{x}]\cdot\bm{\nabla}\psi(\bm{x})+\bm{\nabla}\cdot(\mathsf{D}\bm{\nabla}\psi(\bm{x}))+\left\{\frac{\bm{x}^{\top}\mathsf{V}^{-1}\mathsf{D}\mathsf{V}^{-1}\bm{x}}{4}-\frac{\bm{x}^{\top}\mathsf{V}^{-1}\mathsf{K}\bm{x}}{2}-\frac{\mathrm{tr}(\mathsf{D}\mathsf{V}^{-1})}{2}+\mathrm{tr}(\mathsf{K})\right\}\psi(\bm{x})\notag\\
    &=[(\mathsf{K}-\mathsf{D}\mathsf{V}^{-1})\bm{x}]\cdot\bm{\nabla}\psi(\bm{x})+\bm{\nabla}\cdot(\mathsf{D}\bm{\nabla}\psi(\bm{x}))+\left\{\frac{\mathrm{tr}(\mathsf{K})}{2}-\frac{\bm{x}^{\top}\mathsf{V}^{-1}\mathsf{D}\mathsf{V}^{-1}\bm{x}}{4}\right\}\psi(\bm{x}).
\end{align}
Here, we also used 
\begin{align}
    \mathrm{tr}(\mathsf{D}\mathsf{V}^{-1})=\mathrm{tr}(\mathsf{V}^{-1/2}\mathsf{D}\mathsf{V}^{-1/2})=\mathrm{tr}(\tilde{\mathsf{K}}_+)=\mathrm{tr}(\mathsf{K}),
\end{align}
and
\begin{align}
    \bm{x}^{\top}\mathsf{V}^{-1}\mathsf{D}\mathsf{V}^{-1}\bm{x}&=\frac{\bm{x}^{\top}\mathsf{V}^{-1}(\mathsf{K}\mathsf{V}+\mathsf{V}\mathsf{K}^{\top})\mathsf{V}^{-1}\bm{x}}{2}\notag\\
    &=\frac{\bm{x}^{\top}(\mathsf{V}^{-1}\mathsf{K}+\mathsf{K}^{\top}\mathsf{V}^{-1})\bm{x}}{2}\notag\\
    &=\bm{x}^{\top}\mathsf{V}^{-1}\mathsf{K}\bm{x}.
\end{align}
The adjoint of $\tilde{\mathcal{L}}$ with respect to the standard inner product $\langle\phi,\psi\rangle\coloneqq\int d\bm{x}\,\phi(\bm{x})^{\ast}\psi(\bm{x})$ is also given by
\begin{align}
    \tilde{\mathcal{L}}^{\dag}[\psi(\bm{x})]&=-\bm{\nabla}\cdot[\{(\mathsf{K}-\mathsf{D}\mathsf{V}^{-1})\bm{x}\}\psi(\bm{x})]+\bm{\nabla}\cdot(\mathsf{D}\bm{\nabla}\psi(\bm{x}))+\left\{\frac{\mathrm{tr}(\mathsf{K})}{2}-\frac{\bm{x}^{\top}\mathsf{V}^{-1}\mathsf{D}\mathsf{V}^{-1}\bm{x}}{4}\right\}\psi(\bm{x})\notag\\
    &=-[(\mathsf{K}-\mathsf{D}\mathsf{V}^{-1})\bm{x}]\cdot\bm{\nabla}\psi(\bm{x})-[\mathrm{tr}(\mathsf{K}-\mathsf{D}\mathsf{V}^{-1})]\psi(\bm{x})+\bm{\nabla}\cdot(\mathsf{D}\bm{\nabla}\psi(\bm{x}))+\left\{\frac{\mathrm{tr}(\mathsf{K})}{2}-\frac{\bm{x}^{\top}\mathsf{V}^{-1}\mathsf{D}\mathsf{V}^{-1}\bm{x}}{4}\right\}\psi(\bm{x})\notag\\
    &=-[(\mathsf{K}-\mathsf{D}\mathsf{V}^{-1})\bm{x}]\cdot\bm{\nabla}\psi(\bm{x})+\bm{\nabla}\cdot(\mathsf{D}\bm{\nabla}\psi(\bm{x}))+\left\{\frac{\mathrm{tr}(\mathsf{K})}{2}-\frac{\bm{x}^{\top}\mathsf{V}^{-1}\mathsf{D}\mathsf{V}^{-1}\bm{x}}{4}\right\}\psi(\bm{x}),
\end{align}
where we also used 
\begin{align}
    \mathrm{tr}(\mathsf{K}-\mathsf{D}\mathsf{V}^{-1})=\mathrm{tr}(\mathsf{V}^{1/2}(\tilde{\mathsf{K}}-\tilde{\mathsf{D}})\mathsf{V}^{-1/2})=\mathrm{tr}(\tilde{\mathsf{K}}-\tilde{\mathsf{K}}_+)=0.
\end{align}
Using $\tilde{\mathcal{L}}$ and $\tilde{\mathcal{L}}^{\dag}$, we define the Hermitian and skew-Hermitian parts of $\tilde{\mathcal{L}}$ as
\begin{align}
    \tilde{\mathcal{L}}_{\mathrm{H}}[\psi(\bm{x})]&\coloneqq\frac{\tilde{\mathcal{L}}[\psi(\bm{x})]+\tilde{\mathcal{L}}^{\dag}[\psi(\bm{x})]}{2}=\bm{\nabla}\cdot(\mathsf{D}\bm{\nabla}\psi(\bm{x}))+\left\{\frac{\mathrm{tr}(\mathsf{K})}{2}-\frac{\bm{x}^{\top}\mathsf{V}^{-1}\mathsf{D}\mathsf{V}^{-1}\bm{x}}{4}\right\}\psi(\bm{x}),\notag\\
    \tilde{\mathcal{L}}_{\mathrm{S}}[\psi(\bm{x})]&\coloneqq\frac{\tilde{\mathcal{L}}[\psi(\bm{x})]-\tilde{\mathcal{L}}^{\dag}[\psi(\bm{x})]}{2}=[(\mathsf{K}-\mathsf{D}\mathsf{V}^{-1})\bm{x}]\cdot\bm{\nabla}\psi(\bm{x}).
\end{align}

The normality of $\tilde{\mathcal{L}}$, i.e., $[\tilde{\mathcal{L}},\tilde{\mathcal{L}}^{\dag}]=0$, is equivalent to $[\tilde{\mathcal{L}}_{\mathrm{H}}, \tilde{\mathcal{L}}_{\mathrm{S}}]=0$. Since we can calculate this commutator as
\begin{align}
    &\tilde{\mathcal{L}}_{\mathrm{H}}[\tilde{\mathcal{L}}_{\mathrm{S}}[\psi(\bm{x})]]-\tilde{\mathcal{L}}_{\mathrm{S}}[\tilde{\mathcal{L}}_{\mathrm{H}}[\psi(\bm{x})]]\notag\\
    &=2\bm{\nabla}\cdot[(\mathsf{K}-\mathsf{D}\mathsf{V}^{-1})\mathsf{D}\bm{\nabla}\psi(\bm{x})]+\frac{1}{2}\bm{x}^{\top}\mathsf{V}^{-1}\mathsf{D}\mathsf{V}^{-1}(\mathsf{K}-\mathsf{D}\mathsf{V}^{-1})\bm{x}\psi(\bm{x})\notag\\
    &=2\bm{\nabla}\cdot[\mathsf{V}^{1/2}\tilde{\mathsf{K}}_{-}\tilde{\mathsf{K}}_{+}\mathsf{V}^{1/2}\bm{\nabla}\psi(\bm{x})]+\frac{1}{2}\bm{x}^{\top}\mathsf{V}^{-1/2}\tilde{\mathsf{K}}_{+}\tilde{\mathsf{K}}_{-}\mathsf{V}^{-1/2}\bm{x}\psi(\bm{x})\notag\\
    &=\bm{\nabla}\cdot[\mathsf{V}^{1/2}\{\tilde{\mathsf{K}}_{-}\tilde{\mathsf{K}}_{+}+(\tilde{\mathsf{K}}_{-}\tilde{\mathsf{K}}_{+})^{\top}\}\mathsf{V}^{1/2}\bm{\nabla}\psi(\bm{x})]+\frac{1}{4}\bm{x}^{\top}\mathsf{V}^{-1/2}\{\tilde{\mathsf{K}}_{+}\tilde{\mathsf{K}}_{-}+(\tilde{\mathsf{K}}_{+}\tilde{\mathsf{K}}_{-})^{\top}\}\mathsf{V}^{-1/2}\bm{x}\psi(\bm{x}),
\end{align}
the normality of $\tilde{\mathcal{L}}$ is equivalent to $\tilde{\mathsf{K}}_{-}\tilde{\mathsf{K}}_{+}+(\tilde{\mathsf{K}}_{-}\tilde{\mathsf{K}}_{+})^{\top}=\mathsf{0}$ and $\tilde{\mathsf{K}}_{+}\tilde{\mathsf{K}}_{-}+(\tilde{\mathsf{K}}_{+}\tilde{\mathsf{K}}_{-})^{\top}=\mathsf{0}$. Using $\tilde{\mathsf{K}}_{-}^{\top}=-\tilde{\mathsf{K}}_{-}$ and $\tilde{\mathsf{K}}_{+}^{\top}=\tilde{\mathsf{K}}_{+}$, these conditions are equivalent to
\begin{align}
    \tilde{\mathsf{K}}_{-}\tilde{\mathsf{K}}_{+}-\tilde{\mathsf{K}}_{+}\tilde{\mathsf{K}}_{-}=\mathsf{0},
\end{align}
which implies the normality of $\tilde{\mathsf{K}}$.

\section{Derivation of the decay rate in the short-time regime}
Here, we derive the following relation used in Appendix~C:
\begin{align}
    \gamma^{a}_{0}\coloneqq\liminf_{t\to0^{+}}\frac{1}{t}\ln\left|\frac{C^{a}_{0}}{C^{a}_{t}}\right|=\frac{\bm{a}^{\top}\tilde{\mathsf{K}}_{+}\bm{a}}{\bm{a}^{\top}\bm{a}},
    \label{short-time decay rate in Appendix}
\end{align}
where $t\to0^+$ means approaching zero from the right.
Substituting $C^a_t=\bm{a}^{\top}\mathrm{e}^{-\tilde{\mathsf{K}}t}\bm{a}$ into the definition of $\gamma^{a}_{0}$, we obtain
\begin{align}
    \gamma^{a}_{0}&=\liminf_{t\to0^{+}}\frac{1}{t}\ln\left|\frac{\bm{a}^{\top}\bm{a}}{\bm{a}^{\top}\mathrm{e}^{-\tilde{\mathsf{K}}t}\bm{a}}\right|\notag\\
    &=\liminf_{t\to0^{+}}\frac{1}{t}\ln\left|\frac{\bm{a}^{\top}\bm{a}}{\bm{a}^{\top}\bm{a}-t\bm{a}^{\top}\tilde{\mathsf{K}}\bm{a}+O(t^2)}\right|\notag\\
    &=-\liminf_{t\to0^{+}}\frac{1}{t}\ln\left|1-t\frac{\bm{a}^{\top}\tilde{\mathsf{K}}\bm{a}}{\bm{a}^{\top}\bm{a}}+O(t^2)\right|\notag\\
    &=\liminf_{t\to0^{+}}\left\{\frac{\bm{a}^{\top}\tilde{\mathsf{K}}\bm{a}}{\bm{a}^{\top}\bm{a}}+O(t)\right\}\notag\\
    &=\frac{\bm{a}^{\top}\tilde{\mathsf{K}}\bm{a}}{\bm{a}^{\top}\bm{a}}.
    \label{calc short-time decay rate}
\end{align}
Because $\bm{a}^{\top}\tilde{\mathsf{K}}\bm{a}=(\bm{a}^{\top}\tilde{\mathsf{K}}\bm{a})^{\top}=\bm{a}^{\top}\tilde{\mathsf{K}}^{\top}\bm{a}$, we obtain $\bm{a}^{\top}\tilde{\mathsf{K}}\bm{a}=\bm{a}^{\top}(\tilde{\mathsf{K}}+\tilde{\mathsf{K}}^{\top})\bm{a}/2=\bm{a}^{\top}\tilde{\mathsf{K}}_{+}\bm{a}$. Combining this relation and Eq.~\eqref{calc short-time decay rate}, we obtain Eq.~\eqref{short-time decay rate in Appendix}. We note that we can also express $\gamma_0^a$ as
\begin{align}
    \gamma^{a}_{0}=\frac{\bm{a}^{\dag}\tilde{\mathsf{K}}_{+}\bm{a}}{\bm{a}^{\dag}\bm{a}},
    \label{short-time decay rate}
\end{align}
since $\bm{a}$ is real and $\bm{a}^{\top}=\bm{a}^{\dag}$.

\section{The eigenmode expansion of EPR}
We introduce the expression of $\sigma^{\mathrm{st}}$ in terms of the eigenvalues and eigenvectors of $\mathsf{K}_D$ developed in Ref.~\cite{fyodorov2025nonorthogonal} and discuss the relation with our decomposition. For this purpose, we assume that $\mathsf{K}$ is diagonalizable. 

We consider the coordinate system where the diffusion matrix becomes the identity. In this coordinate system, the drift matrix and the covariance matrix are given by $\mathsf{K}_D$ and $\mathsf{V}_D$, respectively. Using the definitions of these matrices, we can rewrite the EPR [Eq.~(3)] as
\begin{align}
    \sigma^{\mathrm{st}}=\mathrm{tr}[\mathsf{V}_D(\mathsf{K}_D-\mathsf{V}_D^{-1})^{\top}(\mathsf{K}_D-\mathsf{V}_D^{-1})].
\end{align}
In the following, we transform this expression as follows:
\begin{align}
    \sigma^{\mathrm{st}}&=\mathrm{tr}[\mathsf{V}_D(\mathsf{K}_D-\mathsf{V}_D^{-1})^{\top}(\mathsf{K}_D-\mathsf{V}_D^{-1})]\notag\\
    &=\mathrm{tr}[(\mathsf{K}_D-\mathsf{V}_D^{-1})^{\top}(\mathsf{K}_D\mathsf{V}_D-\mathsf{I})]\notag\\
    &=\frac{1}{2}\mathrm{tr}[(\mathsf{K}_D-\mathsf{V}_D^{-1})^{\top}(\mathsf{K}_D\mathsf{V}_D-\mathsf{V}_D\mathsf{K}_D^{\top})]\notag\\
    &=\frac{1}{2}\mathrm{tr}[\mathsf{K}_D^{\top}(\mathsf{K}_D\mathsf{V}_D-\mathsf{V}_D\mathsf{K}_D^{\top})]-\frac{1}{2}\mathrm{tr}[\mathsf{V}_D^{-1}(\mathsf{K}_D\mathsf{V}_D-\mathsf{V}_D\mathsf{K}_D^{\top})]\notag\\
    &=\frac{1}{2}\mathrm{tr}[\mathsf{K}_D^{\top}(\mathsf{K}_D\mathsf{V}_D-\mathsf{V}_D\mathsf{K}_D^{\top})].
    \label{EPR transform}
\end{align}
Here, we used the cyclicity of trace to obtain the second line. The third line follows from the Lyapunov equation $\mathsf{K}_D\mathsf{V}_D+\mathsf{V}_D\mathsf{K}_D^{\top}-2\mathsf{I}=\mathsf{0}$ as
\begin{align}
    \mathsf{K}_D\mathsf{V}_D-\mathsf{I}=\mathsf{K}_D\mathsf{V}_D-\frac{1}{2}(\mathsf{K}_D\mathsf{V}_D+\mathsf{V}_D\mathsf{K}_D^{\top})=\frac{1}{2}(\mathsf{K}_D\mathsf{V}_D-\mathsf{V}_D\mathsf{K}_D^{\top}).
\end{align}
The last line of Eq.~\eqref{EPR transform} is obtained by
\begin{align}
    \mathrm{tr}[\mathsf{V}_D^{-1}(\mathsf{K}_D\mathsf{V}_D-\mathsf{V}_D\mathsf{K}_D^{\top})]=\mathrm{tr}(\mathsf{V}_D^{-1}\mathsf{K}_D\mathsf{V}_D)-\mathrm{tr}(\mathsf{K}_D^{\top})=\mathrm{tr}(\mathsf{K}_D)-\mathrm{tr}(\mathsf{K}_D^{\top})=0.
\end{align}
We can further transform the EPR as
\begin{align}
    \sigma^{\mathrm{st}}&=-\frac{1}{2}\mathrm{tr}[(\mathsf{K}_D\mathsf{V}_D-\mathsf{V}_D\mathsf{K}_D^{\top})\mathsf{K}_D]\notag\\
    &=-\frac{1}{2}\mathrm{tr}[(\mathsf{K}_D\mathsf{V}_D-\mathsf{V}_D\mathsf{K}_D^{\dag})\mathsf{K}_D],
    \label{EPR transform 2}
\end{align}
where we used $\mathrm{tr}[\mathsf{K}_D^{\top}(\mathsf{K}_D\mathsf{V}_D-\mathsf{V}_D\mathsf{K}_D^{\top})]=\mathrm{tr}[\{\mathsf{K}_D^{\top}(\mathsf{K}_D\mathsf{V}_D-\mathsf{V}_D\mathsf{K}_D^{\top})\}^{\top}]=-\mathrm{tr}[(\mathsf{K}_D\mathsf{V}_D-\mathsf{V}_D\mathsf{K}_D^{\top})\mathsf{K}_D]$ in the first transformation. The second transformation follows from the fact that $\mathsf{K}_D$ is real. Using the cyclicity of trace, we obtain
\begin{align}
    \sigma^{\mathrm{st}}=-\frac{1}{2}\mathrm{tr}[\mathsf{K}_D(\mathsf{K}_D\mathsf{V}_D-\mathsf{V}_D\mathsf{K}_D^{\dag})].
    \label{EPR asymmetry}
\end{align}

Based on Eq.~\eqref{EPR asymmetry}, we express the EPR in terms of the eigenvalues and eigenvectors of $\mathsf{K}_D$. Since $\mathsf{K}$ is diagonalizable, we can diagonalize $\mathsf{K}_D=\mathsf{D}^{-1/2}\mathsf{K}\mathsf{D}^{1/2}$ as
\begin{align}
    \mathsf{K}_D=\sum_{n=1}^{N}\lambda_n\bm{r}_n\bm{l}_n^{\dag},
    \label{spectral decomp K_D}
\end{align}
where $\bm{r}_n$ and $\bm{l}_n^{\dag}$ are right and left eigenvectors of $\mathsf{K}_D$ corresponding to $\lambda_n$. These eigenvectors are biorthogonalized so that  $\bm{l}_n^{\dag}\bm{r}_m=\delta_{nm}$ is satisfied. Using this spectral decomposition of $\mathsf{K}_D$, we can rewrite the covariance matrix $\mathsf{V}_D=2\int_{0}^{\infty}dt\,\mathrm{e}^{-\mathsf{K}_Dt}\mathrm{e}^{-\mathsf{K}_D^{\top}t}$ as
\begin{align}
    \mathsf{V}_D&=2\int_{0}^{\infty}dt\,\mathrm{e}^{-\mathsf{K}_Dt}\mathrm{e}^{-\mathsf{K}_D^{\dag}t}\notag\\
    &=2\int_{0}^{\infty}dt\,\sum_{n=1}^{N}\sum_{m=1}^{N}\mathrm{e}^{-(\lambda_n+\lambda_m^{\ast}) t}\bm{r}_n\bm{l}_n^{\dag}\bm{l}_m\bm{r}_m^{\dag}\notag\\
    &=\sum_{n=1}^{N}\sum_{m=1}^{N}\frac{2}{\lambda_n+\lambda_m^{\ast}} (\bm{l}_n^{\dag}\bm{l}_m)\bm{r}_n\bm{r}_m^{\dag}.
    \label{spectral decomp V_D}
\end{align}
Here we used the fact that $\mathsf{K}_D$ is real in the first line.
We also used $\mathrm{Re}(\lambda_n+\lambda_m^{\ast})=\mathrm{Re}(\lambda_n)+\mathrm{Re}(\lambda_m)>0$ in the last transformation. We substitute Eqs.~\eqref{spectral decomp K_D} and ~\eqref{spectral decomp V_D} into Eq.~\eqref{EPR asymmetry}. First, we consider $\mathsf{K}_D\mathsf{V}_D-\mathsf{V}_D\mathsf{K}_D^{\dag}$ in the trace. This term is computed as
\begin{align}
    \mathsf{K}_D\mathsf{V}_D-\mathsf{V}_D\mathsf{K}_D^{\dag}&=\sum_{k=1}^N\sum_{n=1}^N\sum_{m=1}^N\left[\frac{2\lambda_k}{\lambda_n+\lambda_m^{\ast}}(\bm{l}_n^{\dag}\bm{l}_m)\bm{r}_k\bm{l}_k^{\dag}\bm{r}_n\bm{r}_m^{\dag}-\frac{2\lambda_k^{\ast}}{\lambda_n+\lambda_m^{\ast}}(\bm{l}_n^{\dag}\bm{l}_m)\bm{r}_n\bm{r}_m^{\dag}\bm{l}_k\bm{r}_k^{\dag}\right]\notag\\
    &=\sum_{k=1}^N\sum_{n=1}^N\sum_{m=1}^N\left[\frac{2\lambda_k}{\lambda_n+\lambda_m^{\ast}}(\bm{l}_n^{\dag}\bm{l}_m)\bm{r}_k\delta_{kn}\bm{r}_m^{\dag}-\frac{2\lambda_k^{\ast}}{\lambda_n+\lambda_m^{\ast}}(\bm{l}_n^{\dag}\bm{l}_m)\bm{r}_n\delta_{km}\bm{r}_k^{\dag}\right]\notag\\
    &=\sum_{n=1}^N\sum_{m=1}^N\left[\frac{2\lambda_n}{\lambda_n+\lambda_m^{\ast}}(\bm{l}_n^{\dag}\bm{l}_m)\bm{r}_n\bm{r}_m^{\dag}-\frac{2\lambda_m^{\ast}}{\lambda_n+\lambda_m^{\ast}}(\bm{l}_n^{\dag}\bm{l}_m)\bm{r}_n\bm{r}_m^{\dag}\right]\notag\\
    &=\sum_{n=1}^N\sum_{m=1}^N\frac{2(\lambda_n-\lambda_m^{\ast})}{\lambda_n+\lambda_m^{\ast}}(\bm{l}_n^{\dag}\bm{l}_m)\bm{r}_n\bm{r}_m^{\dag}.
\end{align}
This result leads to
\begin{align}
    \mathsf{K}_D(\mathsf{K}_D\mathsf{V}_D-\mathsf{V}_D\mathsf{K}_D^{\dag})&=\sum_{k=1}^N\sum_{n=1}^N\sum_{m=1}^N\frac{2\lambda_k(\lambda_n-\lambda_m^{\ast})}{\lambda_n+\lambda_m^{\ast}}(\bm{l}_n^{\dag}\bm{l}_m)\bm{r}_k\bm{l}_k^{\dag}\bm{r}_n\bm{r}_m^{\dag}\notag\\
    &=\sum_{n=1}^N\sum_{m=1}^N\frac{2\lambda_n(\lambda_n-\lambda_m^{\ast})}{\lambda_n+\lambda_m^{\ast}}(\bm{l}_n^{\dag}\bm{l}_m)\bm{r}_n\bm{r}_m^{\dag}.
\end{align}
Taking the trace of both sides, we express the EPR as
\begin{align}
    \sigma^{\mathrm{st}}&=-\frac{1}{2}\mathrm{tr}[\mathsf{K}_D(\mathsf{K}_D\mathsf{V}_D-\mathsf{V}_D\mathsf{K}_D^{\dag})]\notag\\
    &=-\mathrm{tr}\left[\sum_{n=1}^N\sum_{m=1}^N\frac{\lambda_n(\lambda_n-\lambda_m^{\ast})}{\lambda_n+\lambda_m^{\ast}}(\bm{l}_n^{\dag}\bm{l}_m)\bm{r}_n\bm{r}_m^{\dag}\right]\notag\\
    &=-\sum_{k=1}^N\sum_{n=1}^N\sum_{m=1}^N\frac{\lambda_n(\lambda_n-\lambda_m^{\ast})}{\lambda_n+\lambda_m^{\ast}}(\bm{l}_n^{\dag}\bm{l}_m)\bm{l}_k^{\dag}\bm{r}_n\bm{r}_m^{\dag}\bm{r}_k\notag\\
    &=-\sum_{n=1}^N\sum_{m=1}^N\frac{\lambda_n(\lambda_n-\lambda_m^{\ast})}{\lambda_n+\lambda_m^{\ast}}(\bm{l}_n^{\dag}\bm{l}_m)(\bm{r}_m^{\dag}\bm{r}_n).
    \label{EPR eigenmodes}
\end{align}

Introducing the overlap matrix $\mathsf{O}=(O_{nm})$ as ${O}_{nm}\coloneqq(\bm {l}_n^{\dag}\bm{l}_m)(\bm{r}_m^{\dag}\bm{r}_n)$~\cite{chalker1998eigenvector}, we can rewrite Eq.~\eqref{EPR eigenmodes} as
\begin{align}
    \sigma^{\mathrm{st}}=-\sum_{n=1}^N\sum_{m=1}^N\frac{\lambda_n(\lambda_n-\lambda_m^{\ast})}{\lambda_n+\lambda_m^{\ast}}O_{nm},
    \label{Fyodorov decomp}
\end{align}
which was originally obtained in Ref.~\cite{fyodorov2025nonorthogonal}.
The overlap matrix is related to the nonnormality of $\mathsf{K}_D$. If $\mathsf{K}_D$ is normal, $\mathsf{K}_D$ is diagonalized by a unitary matrix, and the eigenvectors satisfy $\bm{r}_n=\bm{l}_n$. In this case, the biorthogonality leads to $\bm{r}_n^{\dag}\bm{r}_m=\delta_{nm}$. Due to these relations, the overlap matrix is given by the identity matrix as ${O}_{nm}=(\bm{l}_n^{\dag}\bm{l}_m)(\bm{r}_m^{\dag}\bm{r}_n)=(\bm{r}_n^{\dag}\bm{r}_m)(\bm{r}_m^{\dag}\bm{r}_n)=\delta_{nm}$. Thus, we can regard $\mathsf{O}-\mathsf{I}$ as a measure of the nonnormality as follows: The off-diagonal elements $(\mathsf{O}-\mathsf{I})_{nm}={O}_{nm}$ for $n\neq m$ appear due to the nonnormality; the diagonal elements $(\mathsf{O}-\mathsf{I})_{nn}={O}_{nn}-1$ become positive due to the nonnormality as
\begin{align}
    {O}_{nn}-1=(\bm{l}_n^{\dag}\bm{l}_n)(\bm{r}_n^{\dag}\bm{r}_n)-1\geq|\bm{l}_n^{\dag}\bm{r}_n|^2-1=0,
\end{align}
where we used the Cauchy--Schwarz inequality. 
Based on this property of the overlap matrix, we can relate Eq.~\eqref{Fyodorov decomp} to our decomposition. We first rewrite Eq.~\eqref{Fyodorov decomp} as
\begin{align}
    \sigma^{\mathrm{st}}=-\sum_{n=1}^N\frac{\lambda_n(\lambda_n-\lambda_n^{\ast})}{\lambda_n+\lambda_n^{\ast}}-\sum_{n=1}^N\sum_{m=1}^N\frac{\lambda_n(\lambda_n-\lambda_m^{\ast})}{\lambda_n+\lambda_m^{\ast}}(\mathsf{O}-\mathsf{I})_{nm}.
    \label{Fyodorov decomp 2}
\end{align}
The first term represents the dissipation due to the eigenvalues. Indeed, using $\lambda_n-\lambda_n^{\ast}=2\mathrm{i}\mathrm{Im}(\lambda_n)$ and $\lambda_n+\lambda_n^{\ast}=2\mathrm{Re}(\lambda_n)$, we can show that this term equals $\sigma^{\mathrm{osc}}$ as
\begin{align}
    -\sum_{n=1}^N\frac{\lambda_n(\lambda_n-\lambda_n^{\ast})}{\lambda_n+\lambda_n^{\ast}}&=\sum_{n=1}^N\left(-\mathrm{i}\mathrm{Im}(\lambda_n)+\frac{\mathrm{Im}(\lambda_n)^2}{\mathrm{Re}(\lambda_n)}\right)=\sum_{n=1}^N\frac{\mathrm{Im}(\lambda_n)^2}{\mathrm{Re}(\lambda_n)}=\sigma^{\mathrm{osc}}.
\end{align}
Here, in the second transformation, we used $\sum_{n=1}^{N}\mathrm{Im}(\lambda_n)=0$, which is obtained as follows: Since $\mathsf{K}$ is a real matrix, its complex eigenvalues occur in conjugate pairs. Consequently, the sum of their imaginary parts vanishes due to the cancellation between $\mathrm{Im}(\lambda_n)$ and $\mathrm{Im}(\lambda_n^{\ast}) = -\mathrm{Im}(\lambda_n)$. We can also regard the second term in Eq.~\eqref{Fyodorov decomp 2} as $\sigma^{\mathrm{nn}}$, i.e.,
\begin{align}
    \sigma^{\mathrm{nn}}&=-\sum_{n=1}^N\sum_{m=1}^N\frac{\lambda_n(\lambda_n-\lambda_m^{\ast})}{\lambda_n+\lambda_m^{\ast}}(\mathsf{O}-\mathsf{I})_{nm}\notag\\
    &=\sum_{n=1}^N\left[\frac{\mathrm{Im}(\lambda_n)^2}{\mathrm{Re}(\lambda_n)}-\mathrm{i}\mathrm{Im}(\lambda_n)\right]({O}_{nn}-1)-\sum_{n=1}^N\sum_{\substack{m=1\\(m\neq n)}}^N\frac{\lambda_n(\lambda_n-\lambda_m^{\ast})}{\lambda_n+\lambda_m^{\ast}}{O}_{nm}\notag\\
    &=\sum_{n=1}^N\frac{\mathrm{Im}(\lambda_n)^2}{\mathrm{Re}(\lambda_n)}({O}_{nn}-1)-\sum_{n=1}^N\sum_{\substack{m=1\\(m\neq n)}}^N\frac{\lambda_n(\lambda_n-\lambda_m^{\ast})}{\lambda_n+\lambda_m^{\ast}}{O}_{nm},
    \label{nonnormal part Fyodorov}
\end{align}
where we extracted the terms that satisfy $n=m$ to obtain the second line. To obtain the last line, we also used $\sum_{n=1}^N\mathrm{Im}(\lambda_n)({O}_{nn}-1)=0$, which is obtained as follows: Let $\bar{n}$ denote the label of the eigenvalue $\lambda_n^{\ast}$, so that $\lambda_{\bar{n}}=\lambda_n^{\ast}$. Taking the complex conjugate of $\mathsf{K}_D\bm{r}_n=\lambda_n\bm{r}_n$ and $\bm{l}_n^{\dag}\mathsf{K}_D=\lambda_n\bm{l}_n^{\dag}$, we can easily confirm $\bm{r}_{\bar{n}}=\bm{r}_n^{\ast}$ and $\bm{l}_{\bar{n}}=\bm{l}_n^{\ast}$, which yield $O_{\bar{n}\bar{n}}=(\bm{l}_{\bar{n}}^{\dag}\bm{l}_{\bar{n}})(\bm{r}_{\bar{n}}^{\dag}\bm{r}_{\bar{n}})=(\bm{l}_{n}^{\dag}\bm{l}_{n})^{\ast}(\bm{r}_{n}^{\dag}\bm{r}_{n})^{\ast}=(\bm{l}_{n}^{\dag}\bm{l}_{n})(\bm{r}_{n}^{\dag}\bm{r}_{n})=O_{nn}$. This and $\mathrm{Im}(\lambda_{\bar{n}})=\mathrm{Im}(\lambda_{n}^{\ast})=-\mathrm{Im}(\lambda_n)$ imply the desired relation. Since $O_{nn}-1\geq0$, the first term of the last line in Eq.~\eqref{nonnormal part Fyodorov} is nonnegative. 
This term implies that the nonnormality of $\mathsf{K}_D$ enhances the contribution of each complex eigenvalue. 
The second term also implies that the nonnormality induces the contribution from coupling of different eigenmodes. 
Although this expression [Eq.~\eqref{nonnormal part Fyodorov}] is physically easy to interpret, it is not obvious from this expression that $\sigma^{\mathrm{nn}}$ is nonnegative, i.e., the nonnormality increases dissipation. 
This is because each summand in the last term in Eq.~\eqref{nonnormal part Fyodorov} can be complex-valued. In our derivation of the decomposition, we avoid this difficulty due to the complex contributions by applying the Schur decomposition instead of the spectral decomposition. The Schur decomposition also enables us to treat systems with nondiagonalizable drift matrices.

\section{Relation with the mode decomposition of EPR}
We introduce the mode decomposition of the EPR established in Ref.~\cite{sekizawa2024decomposing} using $\tilde{\mathsf{K}}=\mathsf{V}^{-1/2}\mathsf{K}\mathsf{V}^{1/2}$.
In our notation, the mode decomposition of $\sigma^{\mathrm{st}}$ is given as follows.
Since $\tilde{\mathsf{K}}_{-}$ is antisymmetric, its eigenvalues are pure imaginary or zero. This matrix is diagonalized with a unitary basis as $\tilde{\mathsf{K}}_{-}=\sum_{n=1}^{N}\mathrm{i}\Omega_{n}\bm{w}_n\bm{w}_{n}^{\dag}$,
where $\mathrm{i}\Omega_{n}$, with real $\Omega_n$, denotes the $n$th eigenvalue of $\tilde{\mathsf{K}}_{-}$. Here, the right eigenvector corresponding to $\mathrm{i}\Omega_{n}$ is denoted by $\bm{w}_n$, where the eigenvectors are orthonormal as $\bm{w}_n^{\dag}\bm{w}_m=\delta_{nm}$. We rewrite $\sigma^{\mathrm{st}}=\mathrm{tr}(\tilde{\mathsf{K}}_{-}^{\top}\tilde{\mathsf{K}}_{+}^{-1}\tilde{\mathsf{K}}_{-})$ [Eq.~(3)] as $\sigma^{\mathrm{st}}=\mathrm{tr}(\tilde{\mathsf{K}}_{-}^{\dag}\tilde{\mathsf{K}}_{+}^{-1}\tilde{\mathsf{K}}_{-})$ using the fact that $\tilde{\mathsf{K}}_{-}$ is real and satisfies $\tilde{\mathsf{K}}_{-}^{\top}=\tilde{\mathsf{K}}_{-}^{\dag}$. Substituting the spectral decomposition of $\tilde{\mathsf{K}}_{-}$ into this expression of $\sigma^{\mathrm{st}}$, we obtain
\begin{align}
    \sigma^{\mathrm{st}}&=\sum_{n=1}^{N}\sum_{m=1}^{N}\mathrm{tr}[(-\mathrm{i}\Omega_{n}\bm{w}_n\bm{w}_{n}^{\dag})\tilde{\mathsf{K}}_{+}^{-1}(\mathrm{i}\Omega_{m}\bm{w}_m\bm{w}_{m}^{\dag})]\notag\\
    &=\sum_{n=1}^{N}\sum_{m=1}^{N}\Omega_n\Omega_m\mathrm{tr}(\bm{w}_n\bm{w}_{n}^{\dag}\tilde{\mathsf{K}}_{+}^{-1}\bm{w}_m\bm{w}_{m}^{\dag})\notag\\
    &=\sum_{l=1}^{N}\sum_{n=1}^{N}\sum_{m=1}^{N}\Omega_n\Omega_m\bm{w}_l^{\dag}\bm{w}_n\bm{w}_{n}^{\dag}\tilde{\mathsf{K}}_{+}^{-1}\bm{w}_m\bm{w}_{m}^{\dag}\bm{w}_l.
\end{align}
Because $\bm{w}_{l}^{\dag}\bm{w}_n=\delta_{ln}$ and $\bm{w}_{m}^{\dag}\bm{w}_l=\delta_{ml}$ hold, the sum over $l$ and $m$ leaves only the terms that correspond to $l=n$ and $m=n$. Thus, we obtain the mode decomposition as 
\begin{align}
    \sigma^{\mathrm{st}}=\sum_{n=1}^{N}\Omega_n^2\bm{w}_n^{\dag}\tilde{\mathsf{K}}_{+}^{-1}\bm{w}_n.
    \label{mode decomposition}
\end{align}
Here, each summand on the right-hand side is nonnegative due to the positive definiteness of $\tilde{\mathsf{K}}_{+}$. This nonnegativity enables us to regard $\Omega_n^2\bm{w}_n^{\dag}\tilde{\mathsf{K}}_{+}^{-1}\bm{w}_n\geq0$ as the dissipation due to the $n$th eigenmode of $\tilde{\mathsf{K}}_{-}$.

We explain the physical meaning of this mode decomposition.
In the steady state, the flux field is given by $\bm{j}^{\mathrm{st}}(\bm{x})=-p^{\mathrm{st}}(\bm{x})(\mathsf{K}-\mathsf{D}\mathsf{V}^{-1})\bm{x}$. The circulation of this flux field may be characterized by the eigenvalues of $\mathsf{K}-\mathsf{D}\mathsf{V}^{-1}$. These eigenvalues are given by $\{\mathrm{i}\Omega_n\}_{n=1}^{N}$ because $\mathsf{K}-\mathsf{D}\mathsf{V}^{-1}$ is similar to $\tilde{\mathsf{K}}_{-}$. This fact is verified by $\tilde{\mathsf{K}}_{-}=\mathsf{V}^{-{1}/{2}}(\mathsf{K}\mathsf{V}-\mathsf{D})\mathsf{V}^{-{1}/{2}}=\mathsf{V}^{-{1}/{2}}(\mathsf{K}-\mathsf{D}\mathsf{V}^{-1})\mathsf{V}^{{1}/{2}}$.
Thus, we can regard $|\Omega_n|$ as the angular frequency of each oscillatory mode appearing in the flux field; $\Omega_n^2\bm{w}_n^{\dag}\tilde{\mathsf{K}}_{+}^{-1}\bm{w}_n$ is the dissipation due to the corresponding oscillatory mode in the steady state.

If $\tilde{\mathsf{K}}$ is normal, the matrices $\tilde{\mathsf{K}}$, $\tilde{\mathsf{K}}_{+}$, and $\tilde{\mathsf{K}}_{-}$ are diagonalizable with the same basis. In this case, we obtain $\Omega_n=\mathrm{Im}(\lambda_n)$ and $\bm{w}_{n}^{\dag}\tilde{\mathsf{K}}_{+}^{-1}\bm{w}_n=\mathrm{Re}(\lambda_n)^{-1}$.
These relations rewrite each contribution in the mode decomposition~\eqref{mode decomposition} as $\Omega_n^2\bm{w}_n^{\dag}\tilde{\mathsf{K}}_{+}^{-1}\bm{w}_n=\mathrm{Im}(\lambda_n)^2/\mathrm{Re}(\lambda_n)$. Then, the mode decomposition reduces to
\begin{align}
    \sigma^{\mathrm{st}}=\sum_{n=1}^N\frac{\mathrm{Im}(\lambda_n)^2}{\mathrm{Re}(\lambda_n)},
\end{align}
which is consistent with $\sigma^{\mathrm{st}}=\sigma^{\mathrm{osc}}$ and Eq.~(7).
If $\tilde{\mathsf{K}}$ is nonnormal, such a clear relationship between mode decomposition and the eigenvalues of $\mathsf{K}$ is lost, since $\sum_{n=1}^N\Omega_n^2\bm{w}_n^{\dag}\tilde{\mathsf{K}}_{+}^{-1}\bm{w}_n$ also includes the dissipation due to the nonnormality $\sigma^{\mathrm{nn}}$.

\section{Additive reversibilization of a Fokker--Planck generator}
We consider the correspondence of additive reversibilization for the Fokker--Planck generator. First, we explain additive reversibilization for Markov jump processes with discrete states. Let $\partial_t P_i = \sum_j R_{ij}P_j$ be the master equation for Markov jump processes, and $P^{\mathrm{st}}_i$ be the steady-state distribution that satisfies $ \sum_j R_{ij}P^{\mathrm{st}}_j=0$ for any $i$. For Markov jump processes with discrete states, additive reversibilization of the rate  $R_{ij}$~\cite{fill1991eigenvalue,bremaud2013markov,sakai2016eigenvalue,kolchinsky2024thermodynamic} is introduced as $R^{\mathrm{eq}}_{ij} \coloneqq (R_{ij} +P^{\mathrm{st}}_i R_{ji} (P^{\mathrm{st}}_j)^{-1} )/2$. This rate $R^{\mathrm{eq}}_{ij}$ satisfies the following properties: $\sum_j R^{\mathrm{eq}}_{ij}P^{\mathrm{st}}_j=0$,  $R^{\mathrm{eq}}_{ij}P^{\mathrm{st}}_j=R^{\mathrm{eq}}_{ji}P^{\mathrm{st}}_i$, $R^{\mathrm{eq}}_{ii}=R_{ii}$, and $R^{\mathrm{eq}}_{ij}P^{\mathrm{st}}_j+R^{\mathrm{eq}}_{ji}P^{\mathrm{st}}_i = R_{ij}P^{\mathrm{st}}_j+R_{ji}P^{\mathrm{st}}_i$ for any $(i,j)$. Thus, $R^{\mathrm{eq}}_{ij}$ is regarded as the reference rate, which is the equilibrium analogue of $R_{ij}$. Note that the detailed balance condition $R^{\mathrm{eq}}_{ij}P^{\mathrm{st}}_j=R^{\mathrm{eq}}_{ji}P^{\mathrm{st}}_i$ is rewritten as $(P^{\mathrm{st}}_i)^{-1/2} R^{\mathrm{eq}}_{ij}(P^{\mathrm{st}}_j)^{1/2}= (P^{\mathrm{st}}_j)^{-1/2} R^{\mathrm{eq}}_{ji}(P^{\mathrm{st}}_i)^{1/2}$.

Similarly, for the general Fokker--Planck generator $\mathcal{L}_{\mathrm{FP}}$ for continuous-state Markov processes, defined by $\partial_t p_t(\bm{x}) =\mathcal{L}_{\mathrm{FP}}[p_t(\bm{x})]$, we can introduce additive reversibilization as
\begin{align}
\mathcal{L}^{\mathrm{eq}}_{\mathrm{FP}}[p(\bm{x})]\coloneqq\frac{\mathcal{L}_{\mathrm{FP}}[p(\bm{x})]+ p^{\mathrm{st}}(\bm{x})\mathcal{L}^{\dag}_{\mathrm{FP}}[(p^{\mathrm{st}}(\bm{x}))^{-1}p(\bm{x})]}{2},
\label{FPaddtive}
\end{align}
where $p^{\mathrm{st}}(\bm{x})$ is the steady-state distribution that satisfies $\mathcal{L}_{\mathrm{FP}}[p^{\mathrm{st}}(\bm{x})]=0$, and $\mathcal{L}_{\mathrm{FP}}^{\dag}$ is the adjoint of $\mathcal{L}_{\mathrm{FP}}$ with respect to the standard inner product $\langle\phi,\psi\rangle\coloneqq\int d\bm{x}\,\phi(\bm{x})^{\ast}\psi(\bm{x})$. We can confirm that $\mathcal{L}^{\mathrm{eq}}_{\mathrm{FP}}[p^{\mathrm{st}}(\bm{x})]=0$ is satisfied. Moreover, considering the transformation, 
\begin{align}
\tilde{\mathcal{L}}^{\mathrm{eq}}_{\mathrm{FP}}[\psi(\bm{x})]\coloneqq(p^{\mathrm{st}}(\bm{x}))^{-\frac{1}{2}} \mathcal{L}^{\mathrm{eq}}_{\mathrm{FP}}[(p^{\mathrm{st}}(\bm{x}))^{\frac{1}{2}}\psi(\bm{x})],
\end{align}
we obtain the self-adjointness of $\tilde{\mathcal{L}}^{\mathrm{eq}}_{\mathrm{FP}}$,
\begin{align}
\tilde{\mathcal{L}}^{\mathrm{eq}}_{\mathrm{FP}}=(\tilde{\mathcal{L}}^{\mathrm{eq}}_{\mathrm{FP}})^{\dag},
\end{align}
which corresponds to the detailed balance condition $(P^{\mathrm{st}}_i)^{-1/2} R^{\mathrm{eq}}_{ij}(P^{\mathrm{st}}_j)^{1/2}= (P^{\mathrm{st}}_j)^{-1/2} R^{\mathrm{eq}}_{ji}(P^{\mathrm{st}}_i)^{1/2}$ for the reference rate $R^{\mathrm{eq}}_{ij}$. Note that this Fokker--Planck analogue [Eq.~\eqref{FPaddtive}] is introduced as a symmetric operator in Ref.~\cite{qian2013decomposition}.

For OU processes $\mathcal{L}_{\mathrm{FP}}=\mathcal{L}$, this additive reversibilization provides Eq.~(12), i.e., $\mathsf{K}_{\mathrm{eq}} = (\mathsf{K}+\mathsf{V}\mathsf{K}^{\top}\mathsf{V}^{-1})/2$. Here, we show that $\mathsf{K}_{\mathrm{eq}}$ can be defined by the linear force $-\mathsf{K}_{\mathrm{eq}} \bm{x}$ for the reference generator $\mathcal{L}^{\mathrm{eq}}[p(\bm{x})]\coloneqq(\mathcal{L}[p(\bm{x})]+ p^{\mathrm{st}}(\bm{x})\mathcal{L}^{\dag}[(p^{\mathrm{st}}(\bm{x}))^{-1}p(\bm{x})])/2$ as follows: 
\begin{align}
\mathcal{L}^{\mathrm{eq}}[p(\bm{x})] &=\bm{\nabla}\cdot[p(\bm{x}) \mathsf{K}_{\mathrm{eq}}\bm{x} + \mathsf{D} \bm{\nabla} p(\bm{x})].
\label{def.L^eq}
\end{align}
Using $p^{\mathrm{st}}(\bm{x})\propto\exp(-\bm{x}^{\top}\mathsf{V}^{-1}\bm{x}/2)$, $\mathcal{L}^{\dag}[q(\bm{x})]= -(\mathsf{K}\bm{x})\cdot\bm{\nabla}q(\bm{x}) + \bm{\nabla} \cdot [\mathsf{D} \bm{\nabla} q(\bm{x})]$, $\mathsf{K}\mathsf{V} +\mathsf{V}\mathsf{K}^{\top}-2\mathsf{D}=\mathsf{0}$ and $\mathrm{tr}(\mathsf{K})=\mathrm{tr}(\mathsf{V}\mathsf{K}^{\top}\mathsf{V}^{-1})$, we obtain
\begin{align}
&\mathcal{L}^{\mathrm{eq}}[p(\bm{x})]\notag\\
=&\frac{\mathcal{L}[p(\bm{x})]+ p^{\mathrm{st}}(\bm{x})\mathcal{L}^{\dag}[(p^{\mathrm{st}}(\bm{x}))^{-1}p(\bm{x})]}{2}\notag\\
=& \frac{\bm{\nabla}\cdot[p(\bm{x})\mathsf{K}\bm{x}]+\bm{\nabla}\cdot[\mathsf{D}\bm{\nabla} p(\bm{x})]- p^{\mathrm{st}}(\bm{x})(\mathsf{K}\bm{x})\cdot\bm{\nabla}[(p^{\mathrm{st}}(\bm{x}))^{-1}p(\bm{x})]+ p^{\mathrm{st}}(\bm{x})\bm{\nabla}\cdot[\mathsf{D}\bm{\nabla}( (p^{\mathrm{st}}(\bm{x}))^{-1}p(\bm{x}))]}{2} \notag\\
=&\frac{\bm{\nabla}\cdot [p(\bm{x})\mathsf{K}\bm{x}] -(\mathsf{K}\bm{x}) \cdot\bm{\nabla}p(\bm{x})- (\mathsf{K}\bm{x}) \cdot (\mathsf{V}^{-1}\bm{x})p(\bm{x}) +  \bm{\nabla}\cdot[\mathsf{D}\mathsf{V}^{-1}\bm{x}p(\bm{x})]+ \bm{x}^{\top}\mathsf{V}^{-1}\mathsf{D} \mathsf{V}^{-1}\bm{x}p(\bm{x}) + (\mathsf{D}  \mathsf{V}^{-1} \bm{x} )
\cdot \bm{\nabla} p(\bm{x})}{2} \notag\\
&+ \bm{\nabla} \cdot \mathsf{D} \bm{\nabla} p(\bm{x}) \notag\\
=& \frac{\bm{\nabla}\cdot[p(\bm{x})\mathsf{K}\bm{x}]-(\mathsf{K}\bm{x})\cdot\bm{\nabla}  p(\bm{x}) +  \bm{\nabla} \cdot [\frac{\mathsf{K}+\mathsf{V}\mathsf{K}^{\top}\mathsf{V}^{-1}}{2} \bm{x}p(\bm{x})]+(\frac{\mathsf{K}+\mathsf{V}\mathsf{K}^{\top} \mathsf{V}^{-1} }{2}  \bm{x} )
\cdot \bm{\nabla} p(\bm{x})}{2} + \bm{\nabla} \cdot \mathsf{D} \bm{\nabla} p(\bm{x}) \notag\\
=& \frac{\mathrm{tr}(\mathsf{K})p(\bm{x}) + \bm{\nabla} \cdot [(\mathsf{K}+\mathsf{V}\mathsf{K}^{\top}\mathsf{V}^{-1}) \bm{x}p(\bm{x})]- \mathrm{tr}(\frac{\mathsf{K}+\mathsf{V}\mathsf{K}^{\top}\mathsf{V}^{-1} }{2}) p(\bm{x})}{2} + \bm{\nabla}\cdot\mathsf{D}\bm{\nabla}p(\bm{x}) \notag\\
=&\bm{\nabla}\cdot\left[p(\bm{x})\frac{\mathsf{K} + \mathsf{V} \mathsf{K}^{\top} \mathsf{V}^{-1}}{2} \bm{x} \right]  + \bm{\nabla} \cdot \mathsf{D} \bm{\nabla} p(\bm{x}).
\label{calc. L^eq}
\end{align}
By comparing Eq.~\eqref{def.L^eq} with Eq.~\eqref{calc. L^eq}, we confirm Eq.~(12). Furthermore, this $\mathsf{K}_{\mathrm{eq}}=(\mathsf{K}+\mathsf{V}\mathsf{K}^{\top} \mathsf{V}^{-1})/2$ is discussed in terms of $\mathsf{K}_{\mathrm{eq}}=\mathsf{D}\mathsf{V}^{-1}$~\cite{lelievre2013optimal,Anton2022}, which has been transformed using the Lyapunov equation $\mathsf{K}\mathsf{V} +\mathsf{V}\mathsf{K}^{\top}-2\mathsf{D}=\mathsf{0}$.

\section{Relationship between the nonnormal EPR and the decay rates in the short-time and long-time regimes}

Here, we consider the difference between the short-time and long-time decay rates of $C_t^a$. We introduce some quantities, which will be used in the following. We define $\gamma^{\star}_{0}$ as the supremum value of $\gamma^a_{0}$,
\begin{align}
    \gamma^{\star}_{0}\coloneqq \sup_{\bm{a} \neq \bm{0}}\gamma^{a}_{0}=\sup_{\bm{a} \neq \bm{0}}\frac{\bm{a}^{\dag}\tilde{\mathsf{K}}_{+}\bm{a}}{\bm{a}^{\dag}\bm{a}},
    \label{Reyleigh-quotient}
\end{align}
where we used Eq.~\eqref{short-time decay rate}.
Because $\gamma^{a}_{0}$ has the Rayleigh-quotient form for the Hermitian matrix $\tilde{\mathsf{K}}_{+}=\tilde{\mathsf{D}}$, maximizing over all $\bm{a} (\neq \bm{0})$ gives the largest eigenvalue of $\tilde{\mathsf{D}}$: 
\begin{align}
\gamma^{\star}_{0} =\lambda_{\mathrm{max}}(\tilde{\mathsf{D}}).
\label{Reyleigh-quotient2}
\end{align}
We will also use the slowest long-time decay rate
\begin{align}
    \gamma_{\infty}\coloneqq\inf_{\bm{a}\neq\bm{0}}\gamma_{\infty}^{a}=\mathrm{Re}(\lambda_1)=\tau_{\mathrm{c}}^{-1}.
\end{align}
Note that $\gamma^{a}_{\infty}=\gamma_{\infty}$ for almost all observables
$a(\bm{x})$, i.e., for generic observables that have nonzero overlap on
the slowest eigenmode. For such observables, the difference between the short-time and long-time decay rates is given by
\begin{align}
    \gamma_{\infty}-\gamma_{0}^a.
\end{align}
In the following, we establish a trade-off between this difference and the nonnormal EPR. 

As a preparation, we introduce the operator norm of a Hermitian $N\times N$ matrix $\mathsf{M}$ as
\begin{align}
    \|\mathsf{M}\|_{\mathrm{op}}\coloneqq\sup_{\bm{u}\neq\bm{0}}\frac{\|\mathsf{M}\bm{u}\|}{\|\bm{u}\|}.
    \label{op norm}
\end{align}
Let $\{\lambda_n(\mathsf{M})\}_{n=1}^N$ denote the eigenvalues of $\mathsf{M}$ repeated according to their algebraic multiplicities, which are labeled so that $|\lambda_1(\mathsf{M})|\leq|\lambda_2(\mathsf{M})|\leq\cdots\leq|\lambda_N(\mathsf{M})|$. Using these eigenvalues, the operator norm of $\mathsf{M}$ is given by
\begin{align}
    \|\mathsf{M}\|_{\mathrm{op}}=|\lambda_N(\mathsf{M})|,
    \label{op norm eigenvalue}
\end{align}
because a Hermitian matrix $M$ is unitarily diagonalizable.
In the following, we focus on $N\geq2$. For any $N\times N$ traceless Hermitian matrix $\mathsf{Z}$, the operator norm provides a lower bound on the Frobenius norm as
\begin{align}
    \|\mathsf{Z}\|_{\mathrm{F}}\geq\sqrt{\frac{N}{N-1}}\|\mathsf{Z}\|_{\mathrm{op}}.
    \label{op norm bound}
\end{align}
This inequality is derived as follows. Since $\mathsf{Z}$ is traceless, we have $\lambda_N(\mathsf{Z})=-\sum_{n=1}^{N-1}\lambda_n(\mathsf{Z})$. Using this and the Cauchy--Schwarz inequality, we obtain $\lambda_N(\mathsf{Z})^2=(\sum_{n=1}^{N-1}\lambda_n(\mathsf{Z}))^2\leq(\sum_{n=1}^{N-1}1^2)(\sum_{n=1}^{N-1}\lambda_n(\mathsf{Z})^2)=(N-1)\sum_{n=1}^{N-1}\lambda_n(\mathsf{Z})^2$. Combining this inequality and $\|\mathsf{Z}\|_{\mathrm{F}}^2=\sum_{n=1}^{N}\lambda_n(\mathsf{Z})^2$, we obtain Eq.~\eqref{op norm bound} as
\begin{align}
    \|\mathsf{Z}\|_{\mathrm{F}}^2&=\lambda_N(\mathsf{Z})^2+\sum_{n=1}^{N-1}\lambda_n(\mathsf{Z})^2\notag\\
    &\geq\lambda_N(\mathsf{Z})^2+\frac{1}{N-1}\lambda_N(\mathsf{Z})^2\notag\\
    &=\frac{N}{N-1}\lambda_N(\mathsf{Z})^2\notag\\
    &=\frac{N}{N-1}\|\mathsf{Z}\|_{\mathrm{op}}^2,
\end{align}
where we also used Eq.~\eqref{op norm eigenvalue}.

Using Eqs.~(21) and ~(23) of Appendix~E in the End Matter, we have 
\begin{align}
\sigma^{\mathrm{nn}}&\geq\frac{1}{\lambda_{\mathrm{max}}(\tilde{\mathsf{D}})}\left\|\mathsf{S}-\mathsf{S}^{\star}\right\|^{2}_{\mathrm{F}}\notag\\
 &\geq\frac{1}{\lambda_{\mathrm{max}}(\tilde{\mathsf{D}})}\left\|\mathsf{H}-\mathsf{H}^{\mathrm{diag}}\right\|^{2}_{\mathrm{F}}.
 \label{ineq in EM}
\end{align}
Because $\mathsf{H}-\mathsf{H}^{\mathrm{diag}}$ is a traceless Hermitian matrix, we can relax this inequality using Eq.~\eqref{op norm bound} as
\begin{align}
\sigma^{\mathrm{nn}}&\geq\frac{1}{\lambda_{\mathrm{max}}(\tilde{\mathsf{D}})}\frac{N}{N-1}\left\|\mathsf{H}-\mathsf{H}^{\mathrm{diag}}\right\|^{2}_{\mathrm{op}}\notag\\
&\geq\frac{1}{\lambda_{\mathrm{max}}(\tilde{\mathsf{D}})}\frac{N}{N-1}\frac{[\bm{b}^{\dagger}(\mathsf{H}^{\mathrm{diag}}-\mathsf{H})\bm{b}]^{2}}{(\bm{b}^{\dag}\bm{b})^2},
\label{nn bound with op norm}
\end{align}
where $\bm{b}=\mathsf{U}^{\dag}\bm{a}$. Here, the second inequality follows from the Cauchy--Schwarz inequality $\|(\mathsf{H}-\mathsf{H}^{\mathrm{diag}})\bm{b}\|\|\bm{b}\|\geq|\bm{b}^{\dag}(\mathsf{H}-\mathsf{H}^{\mathrm{diag}})\bm{b}|$ and Eq.~\eqref{op norm} as $\|\mathsf{H}-\mathsf{H}^{\mathrm{diag}}\|_{\mathrm{op}}\geq\|(\mathsf{H}-\mathsf{H}^{\mathrm{diag}})\bm{b}\|/\|\bm{b}\|\geq|\bm{b}^{\dag}(\mathsf{H}-\mathsf{H}^{\mathrm{diag}})\bm{b}|/(\bm{b}^{\dag}\bm{b})$. We also obtain
\begin{align}
    \frac{\bm{b}^{\dagger}(\mathsf{H}^{\mathrm{diag}}-\mathsf{H})\bm{b}}{\bm{b}^{\dag}\bm{b}}&=\frac{\bm{b}^{\dag}\mathsf{H}^{\mathrm{diag}}\bm{b}}{\bm{b}^{\dag}\bm{b}}-\frac{\bm{b}^{\dag}\mathsf{H}\bm{b}}{\bm{b}^{\dag}\bm{b}}\notag\\
    &\geq\min_{1\leq n\leq N}H_{nn}-\frac{\bm{a}^{\dag}\mathsf{U}\mathsf{H}\mathsf{U}^{\dag}\bm{a}}{\bm{a}^{\dag}\bm{a}}\notag\\
    &=\mathrm{Re}(\lambda_1)-\frac{\bm{a}^{\dag}\tilde{\mathsf{K}}_{+}\bm{a}}{\bm{a}^{\dag}\bm{a}}\notag\\
    &=\gamma_{\infty}-\gamma_0^{a},
    \label{time scale gap bound}
\end{align}
where we used $\tilde{\mathsf{K}}_{+}=\mathsf{U}\mathsf{H}\mathsf{U}^{\dag}$ and Eq.~\eqref{short-time decay rate}. Combining Eqs.~\eqref{Reyleigh-quotient2}, ~\eqref{nn bound with op norm} and ~\eqref{time scale gap bound}, we obtain the trade-off between $\sigma^{\mathrm{nn}}$ and $\gamma_{\infty}-\gamma_0^{a}$ as
\begin{align}
\sigma^{\mathrm{nn}} &\geq \frac{N}{N-1}\frac{[(\gamma_{\infty}-\gamma^{a}_{0})_{+}]^2}{\gamma^{\star}_{0}},
\end{align}
where $(x)_{+}\coloneqq\max\{x,0\}$. This bound implies that the nonnormal EPR is required to make the decay rate of $C_t^a$ in the short-time regime smaller than that in the long-time limit. This effect is consistent with previous results in a random neural model, where nonnormality was found to increase the integrated correlation timescale at a fixed spectrum~\cite{marti2018correlations}. Since the spectrum determines the asymptotic decay rate, this increase reflects the slowing of short-time decay (decreasing of $\gamma_0^a$).

The bound on the relaxation speedup [Eq.~(15)] is recovered by choosing $\bm{a}$ as the eigenvector of $\tilde{\mathsf{K}}_{+}$ corresponding to $\lambda_1^{\mathrm{eq}}$. For this special case, we have $\gamma_0^a=\lambda_1^{\mathrm{eq}}$ and
\begin{align}
\sigma^{\mathrm{nn}} &\geq \frac{N}{N-1}\frac{\{(\mathrm{Re}(\lambda_{1})-\lambda_{1}^{\mathrm{eq}})_{+}\}^2}{\lambda_{\mathrm{max}}(\tilde{\mathsf{D}})}\notag\\
&=\frac{N}{N-1}\frac{[\tau_{\mathrm{c}}^{-1}-(\tau_{\mathrm{c}}^{\mathrm{eq}})^{-1}]^{2}}{\lambda_{\mathrm{max}}(\tilde{\mathsf{D}})},
\end{align}
where we also used $\mathrm{Re}(\lambda_{1})\geq\lambda_1^{\mathrm{eq}}$ and the definitions of the correlation times.

\end{document}

%% file: preamble.tex
\usepackage[utf8]{inputenc}
\usepackage[T1]{fontenc}
\usepackage{newtxtext}
\usepackage{lipsum}

\usepackage{amsmath}
\usepackage{amssymb}
\usepackage{mathtools}
\usepackage{graphicx}
\usepackage{xcolor}
\usepackage{bm}
\usepackage{mathrsfs}
\usepackage{colortbl}
\usepackage{graphicx}
\usepackage[version=4]{mhchem}

\definecolor{myred}{HTML}{E1558A}
\definecolor{myblue}{HTML}{318294}
\definecolor{myskyblue}{HTML}{3FC3E0}

\usepackage[hidelinks]{hyperref}
\hypersetup{
  colorlinks   = true, 
  urlcolor     = blue, 
  linkcolor    = myred, 
  citecolor   = myblue 
}

\DeclareMathOperator*{\argmin}{arg\,min}

%% file: biblio.bbl
%